\documentclass[longauth]{aa}
\usepackage{color}
\usepackage[normalem]{ulem}
\usepackage{comment}

\begin{document}

\date{Received xx / Accepted 8 February 2019} 
  \title{Spatial segregation of dust grains in transition disks\thanks{Based on observations performed with SPHERE/VLT under program ID 099.C-0891(A) and 099.C-0147(A).}}

      \subtitle{SPHERE observations of 2MASS\,J16083070-3828268 and RXJ1852.3-3700}

   \author{M.~Villenave\inst{1,}\inst{2}          
          \and
          M.~Benisty\inst{1,}\inst{3,}\inst{4}
          \and
          W.\,R.\,F.~Dent\inst{5}
          \and
          F.~M\'enard\inst{1}
          \and
          A.~Garufi\inst{6}
          \and
         C.~Ginski\inst{7}
          \and 
          P.~Pinilla\inst{8}
           \and
          C.~Pinte\inst{9,1}
          \and
          J.\,P.~Williams\inst{10}
          \and
          J.~de Boer\inst{11}
          \and
           J.-I. Morino\inst{12}
          \and
           M. Fukagawa\inst{12}
          \and
          C.~Dominik\inst{7}
          \and
          M.~Flock\inst{13}
          \and
          T.~Henning\inst{13}
          \and 
          A.~Juh\'asz\inst{14}
          \and
          M.~Keppler\inst{13}
          \and
          G.~Muro-Arena\inst{7}
          \and
          J.~Olofsson\inst{15,}\inst{16}
          \and
          L.~M.~P\'erez\inst{4} 
          \and
          G.~van der Plas\inst{1}
          \and
          A.~Zurlo\inst{17,}\inst{18,}\inst{19}
          \and
           M.~Carle\inst{19}     
          \and
           P.~Feautrier\inst{1}
          \and
            A.~Pavlov\inst{13}
          \and
            J.~Pragt\inst{21}
          \and
           J.~Ramos\inst{13}
          \and
            J.-F.~Sauvage\inst{20,19}
          \and
           E.~Stadler\inst{1}
          \and
           L.~Weber\inst{22}
    }

   \institute{Univ. Grenoble Alpes, CNRS, IPAG, 38000 Grenoble, France \email{marion.villenave@univ-grenoble-alpes.fr}
              \and 
              European Southern Observatory, Alonso de C\'ordova 3107, Vitacura, Casilla 19001, Santiago 19, Chile
              \and 
              Unidad Mixta Internacional Franco-Chilena de Astronom\'{i}a (CNRS, UMI 3386), 
              \and Departamento de Astronom\'{i}a, Universidad de Chile, Camino El Observatorio 1515, Las Condes, Santiago, Chile
             \and
    		Joint ALMA Observatory, Alonso de C\'ordova 3107, Vitacura 763-0355, Santiago, Chile
    		\and INAF, Osservatorio Astrofisico di Arcetri, Largo Enrico Fermi 5, I-50125 Firenze, Italy
    		\and Anton Pannekoek Institute for Astronomy, University of Amsterdam, Science Park 904,1098XH Amsterdam, The Netherlands
            \and Department of Astronomy/Steward Observatory, The University of Arizona, 933 North Cherry Avenue, Tucson, AZ 85721, USA
            \and Monash Centre for Astrophysics (MoCA) and School of Physics and Astronomy, Monash University, Clayton Vic 3800, Australia
            \and Institute for Astronomy, University of Hawaii, Honolulu, Hawaii, USA
            \and Leiden Observatory, Leiden University, P.O. Box 9513, 2300 RA Leiden, The Netherlands
            \and
    		National Astronomical Observatory of Japan, Osawa 2-21-1, Mitaka, Tokyo, 181-8588, Japan
            \and Max Planck Institute for Astronomy, K\"{o}nigstuhl 17, 69117 Heidelberg, Germany
            \and Institute of Astronomy, Madingley Road, Cambridge CB3 OHA, UK
            \and Facultad de Ciencias, Instituto de F\'isica y Astronom\'ia, Universidad de Valpara\'iso, Av. Gran Breta\~{n}a 1111, 5030 Casilla, Valpara\'iso, Chile
            \and 
            N\'ucleo Milenio Formaci\'on Planetaria - NPF, Universidad de Valpara\'iso, Av. Gran Breta\~na 1111, Valpara\'iso, Chile
            \and N\'ucleo de Astronom\'ia, Facultad de Ingenier\'ia y Ciencias, Universidad Diego Portales, Av. Ejercito 441, Santiago, Chile
            \and Escuela de Ingenier\'ia Industrial, Facultad de Ingenier\'ia y Ciencias, Universidad Diego Portales, Av. Ejercito 441, Santiago, Chile
            \and Aix Marseille Universit\'e, CNRS, LAM, 13388 Marseille, France
    		\and
    		DOTA, ONERA, Universit\'e Paris Saclay, F-91123, Palaiseau France
    		\and
    		NOVA Optical Infrared Instrumentation Group, Oude Hoogeveensedijk 4, 7991 PD Dwingeloo, The Netherlands
    		\and
    		Geneva Observatory, University of Geneva, Chemin des Mailettes 51, 1290 Versoix, Switzerland
		}

 \abstract
 {The mechanisms governing the opening of cavities in transition disks are not fully understood. Several processes have been proposed, but their occurrence rate is still unknown.}
 {We present spatially resolved observations of two transition disks, and aim at constraining their vertical and radial structure using multiwavelength observations that probe different regions of the disks and can help understanding the origin of the cavities.}
 {We have obtained near-infrared scattered light observations with VLT/SPHERE of the transition disks~2MASS~J16083070-3828268~(J1608) and RXJ1852.3-3700 (J1852), located in the Lupus and Corona Australis star-forming regions respectively. We complement our datasets with archival ALMA observations, and with unresolved photometric observations covering a wide range of wavelengths. We performed radiative transfer modeling to analyze the morphology of the disks, and then compare the results with a sample of 20 other transition disks observed with both SPHERE and ALMA.}
 {We detect scattered light in J1608 and J1852 up to a radius of 0.54\arcsec{} and 0.4\arcsec{} respectively. The image of J1608 reveals a very inclined disk (i$\sim$74$^\circ$), with two bright lobes and a large cavity. We also marginally detect the scattering surface from the rear-facing side of the disk. J1852 shows an inner ring extending  beyond the coronagraphic radius up to 15\,au, a gap and a second ring at~42\,au. Our radiative transfer model of J1608 indicates that the  millimeter-sized grains are less extended vertically and radially than the  micron-sized grains, indicating advanced settling and radial drift. We find good agreement with the observations of J1852 with a similar model, but due to the low inclination of the system, the model remains partly degenerate.  The analysis of~22 transition disks shows that, in general, the cavities observed in scattered light are smaller than the ones detected at millimeter wavelengths.}
 {The analysis of a sample of transition disks indicates that the small grains, well coupled to the gas, can flow inward of the region where millimeter grains are trapped.  While 15 out of the 22 cavities in our sample could be explained by a planet of less than 13 Jupiter masses, the others either require the presence of a more massive companion or of several low-mass planets.}

   \maketitle
%

\section{Introduction}

The variety of physical and structural conditions in protoplanetary disks\textemdash the birthplace of planets\textemdash might be responsible for the diversity observed in the exoplanet population. Studying  disk evolution through the analysis of protoplanetary disks with depleted regions and/or clear signs of evolution could provide indirect constraints on the way in which planets form.  Thanks to new capabilities of high resolution instruments such as ALMA, VLT/SPHERE and Gemini/GPI, many features have been identified in protoplanetary disks and in particular in transition disks (TDs). A number of studies reveal rings \citep[e.g.,][]{isella2016,Pohl_2017,Muro-Arena_2018}, lopsided emission \citep[e.g.,][]{casassus2013, cazzoletti2018}, spirals \citep[e.g.,][]{perez2016, dong2018, Muto_2012, Stolker_2017, Benisty_2017, Uyama_2018}, and shadows \citep[e.g.,][]{Marino_2015, Stolker_2017, Casassus_2018, Benisty_2018}. Among other mechanisms,  planets interacting with the disk can form such structures. Although challenging, the observation and/or hints of forming planets within disks have been reported in  recent studies \citep{Keppler_2018, pinte_2018, Teague_2018}.

Interaction with a planetary or stellar companion  is also thought to result in the large cavities observed in circumbinary or transition disks  \citep[e.g.,][]{Munoz_2016, Rosotti_2016, Dipierro_Laibe_2017, Price_2018}. Such disks were identified from a lack of emission in the near-infrared (IR) in their spectral energy distribution (SED), which indicates a dust depleted inner region \citep{Strom_1989}. Other mechanisms such as photo-evaporation~\citep[][]{Owen_2011} or the presence of a dead zone \citep{Flock_2015, Pinilla_2016} are potential processes that could open such cavities. Each mechanism is expected to shape the inner disk differently, and can result in a cavity radius that depends on the dust grain size. However, their respective importance is currently not well constrained. 

While $\mu$m-sized dust are coupled to the gas that orbits at sub-Keplerian speed, larger grains are increasingly decoupled.  Their interactions with the gas lead simultaneously to inward radial drift and vertical settling~\citep{Weidenschilling_1977, Dullemond_2004}. Thus, the large grains are expected to be located in a more compact region radially and to be less extended vertically than smaller grains. However, the strength of these effects is not yet well constrained. Grains of different sizes would also be trapped more or less efficiently by a pressure maximum, which can be generated at the outer edge of dead zones or by a planet in the disk. 

The combination of high-resolution observations at different wavelengths is key for quantifying both the radial and vertical distribution of dust grains, and in particular, the degree of dust settling. It can therefore help to differentiate the various mechanisms that can generate cavities in transition disks.
While direct imaging with SPHERE traces polarized scattered light by small grains~($< \text{few}\,\mu$m) well coupled to the gas and located in the surface layers,  ALMA observations probe thermal emission of larger grains~($>50\,\mu$m), partially decoupled from the gas and located in the midplane. The combination of both tracers therefore allows one to trace and compare different dust grain populations.

In this paper we present scattered light images of two transition disks, \object{2MASS J16083070-3828268} and \object{RXJ1852.3-3700} (hereafter J1608 and J1852, respectively) observed with VLT/SPHERE. We complement our observations with ALMA archival data.  We aim to model both disks using a radiative transfer code and to bring constraints on the radial segregation of dust particles, on vertical settling and on the origin of their cavities.  In Section~\ref{sec:sample}, we present the two transition disks, and in Section~\ref{sec:obs}, the observations and data reduction. The modeling procedure and results are detailed in Section~\ref{sec:model}. We compare our results with a larger sample of transition disks in Section~\ref{sec:discussion}. Finally, the conclusions are presented in Section~\ref{sec:conclusion}.

\section{Stellar and disk properties}
\label{sec:sample}
\begin{table}
\centering
\caption{Stellar parameters.}
\begin{tabular}{|ll|c|c|}
\hline
Parameters && J1608 & J1852\\
\hline
RA &[h m s]&16 08 30.7&18 52 17.3\\
Dec &[deg \arcmin~\arcsec]&-38 28 26.8&-37 00 11.9\\
Distance$^1$ &[pc]&$156 \pm 6$&$146 \pm1$\\
Av&[mag]&0.1&1.0\\

SpT&&K2&K2\\
Teff &[K]&4800&4850\\

R$_\star$ &[R$_\odot$]&2.00&1.17\\
$\dot{M}$&[M$_\odot$.yr$^{-1}$]&10$^{-9}$&10$^{-9}$ \\
M$_\star^{~2}$ &[M$_\odot$]&$1.4 \pm 0.1$&$1.0\pm0.1$\\
Age$^2$ &[Myr] & 8.3$^{+4.4}_{-2.1}$ & >14\\

\hline
\end{tabular}
\tablebib{\citet{alcala_2017, manara_2014} 1/ \citet{gaia_DR2_2018}, 2/ Stellar masses and ages are calculated by \citet{Garufi_2018} using Gaia DR2 distances and stellar tracks by \citet{Siess_2000}.}
\label{tab:stellarParam} 
\end{table}

J1608 and J1852 are two transition disks around K2 stars, located in close-by star-forming regions. 
Using VLT/X-shooter spectroscopy, mass accretion rates of~$\sim$10$^{-9}$~M$_\odot$\,yr$^{-1}$ was found for both objects, typical of transition disks and indicating that the inner disk regions still hold significant gas content  \citep{alcala_2017, manara_2014}. In this paper, we will use the stellar ages and masses re-estimated by \citet{Garufi_2018} using the latest Gaia~DR2 distances \citep{gaia_DR2_2018} and stellar tracks by \citet{Siess_2000}. J1608 is found to be  younger than J1852 although age estimates appear uncertain. We report the stellar parameters in Table~\ref{tab:stellarParam}. All the radial extents provided in physical units in the following have been scaled to the Gaia DR2 distances.

J1608 is located at 156 pc~\citep{gaia_DR2_2018} in the Lupus III cloud. It has been observed with ALMA in Band~7 and Band~6 \citep[$\sim\,0.89$\,mm and~$\sim$\,1.33\,mm, respectively;][]{ansdell_2016, ansdell_2018}. 
J1608 is one of the most massive disks of the Lupus millimeter survey with~$\sim$80\,M$_\oplus$ of dust. It possesses a large cavity in the continuum and is highly inclined~\citep[>70$^\circ$,][]{ansdell_2016}. 
From the Band 6 observations, \citet{ansdell_2018} find a gas radial extent about twice as large as that of the dust.  \citet{Pinilla_2018} modeled the Band~7 continuum emission with a radially asymmetric Gaussian ring model and found that the peak intensity of the ring is located at~$\sim$61\,au. This is consistent with the dust and line modeling of the ALMA data by \citet{van_der_marel_new_2018} that constrains the outer radius of the dust and gas cavity respectively to~59\,au and 47\,au. 

\medskip
J1852 is one of the oldest systems in the Corona Australis association \citep[CrA,][]{neuhaeuser_2000}, located at a distance of 146 pc~\citep{gaia_DR2_2018}.
Gas emission lines of [Ne\,II], [H\,I]  and~[O\,I] were detected~\citep{rigliaco_2015, pascucci_2007, geers_2012, marichalar_2016}, using Spitzer and Herschel data. 
While no spectroscopic binary was found in the system \citep{kohn_2016},  a candidate companion located at~3\arcsec{}~($\sim$\,440\,au) separation, was identified with SUBARU~\citep{uyama_2017}. Follow-up observations are needed to confirm if it is a bound companion or a background object. 

SMA observations, with rather low resolution~(1.0\arcsec{}$\times$1.7\arcsec{}), show that the disk has a relatively low inclination \citep[$\sim$30$^\circ$,][]{hughes_2010}. From  SED modeling, \citet{hughes_2010} inferred the presence of a cavity up to 16\,au,with an optically thick inner disk closer to the star, more recently confirmed by \citet{van_der_marel_whole_2016}. 
\citet{geers_2012} presented a thermo-chemical model explaining the upper limits for the [OI] and CO emission lines, and found that the gas is either optically thin and co-located with the dust (16-500\,au), or possibly radially concentrated in an optically thick region~(16-70\,au).

\medskip
We selected these two transition disks around K2 stars with similar stellar properties, but seen at different inclinations to investigate the differences in their structure and the mechanisms that can be responsible for their inner cavities. 

\section{Observations and data reduction}
\label{sec:obs}
\subsection{Observations}
Both disks were observed with the InfraRed Dual band Imager and Spectrograph \citep[IRDIS,][]{Dohlen_2008} sub-instrument of SPHERE \citep[][]{Beuzit_2008}
mounted on the Very Large Telescope. Observations were carried out in dual-polarization imaging mode~\citep[DPI,][]{Langlois_2014}, in both~J~($\lambda_J=1.245\,\mu$m) and H-band ($\lambda_H=1.625\,\mu$m) for J1608, and in H-band only for J1852. 

We observed J1608, during the nights of June 18, and July 23, 2017 (ID 099.C-0891,~PI:~Benisty). During the first epoch (June 18, 2017), we observed~J1608 in H-band, with an apodized Lyot coronagraphic mask~\citep[N\_ALC\_YJH\_S,~0.185\arcsec{} in diameter;][]{Martinez_2009, Carbillet_2009}. The observations consisted of 56 exposures of 32 seconds each, corresponding to about~30~minutes on source. Conditions were good, with a seeing between 0.65\arcsec{} and 0.9\arcsec{} during the night. To confirm the presence of a cavity, we also performed non-coronagraphic observations on July 23, 2017, in J-band, with about 80~seconds on source, by exposures of 2 seconds. Weather conditions were relatively poor with a seeing between 1.2\arcsec{} and~1.9\arcsec{} during these observations. %
 
 J1852 was observed as part of the SPHERE guaranteed time observations (GTO) program on May~15,~2017~(ID~099.C-0147, PI:~Beuzit). Observations were performed in H-band, using the same coronagraph as for J1608, and consisted in 12 exposures of~64\,seconds each. The seeing was around 1.2\arcsec{} during the observations. 

\medskip
We reduced the data to generate the total intensity map and~Stokes~Q and U polarized maps, following the approach detailed in \citet[][]{Ginski_2016}. 
The polarized intensity (PI) image is computed from the~Stokes~Q and U components:
\begin{equation}
\mathrm{PI} = \sqrt{Q^2 + U^2}
\label{eq:PI}
\end{equation}
We also define the polar Stokes components Q$_\phi$ and U$_\phi$ as in \citet{Schmid_2006}:
\begin{equation}
\centering
Q_\phi = +Q\cos(2\phi) + U \sin(2\phi)
\label{Qphi}
\end{equation}
\begin{equation}
\centering
U_\phi = -Q\sin(2\phi) + U \cos(2\phi)
\label{Uphi}
\end{equation}
where $\phi = \arctan(\frac{y}{x})$ corresponds to the azimuthal angle as measured north to east with respect to the position of the star (centered in the image). The positive signal in Q$_\phi$ is the polarization in the azimuthal direction (negative signal is in the radial direction), while U$_\phi$ represents the polarization inclined by 45$^\circ$ from this direction. In the case of single scattering events, a photon is expected to be polarized orthogonally to the scattering plane, defined by the light source, the scattering particle and the observer. Thus, in this scenario, all the polarized signal should be included in the Q$_\phi$ component. \citet{Canovas_2015} showed however, that if the disk is too inclined and/or multiple scattering events occurs, the polarization is not necessarily perpendicular to the scattering plane. Then, part of the astrophysical signal will be included in the U$_\phi$ component. This effect was indeed observed in the very inclined~($\sim$69$^\circ$) disk around~T\,Cha, leading to a large~U$_\phi$/Q$_\phi$ peak-to-peak value~\citep{Pohl_2017}.

\begin{figure*}
\centering
\includegraphics[width =18cm]{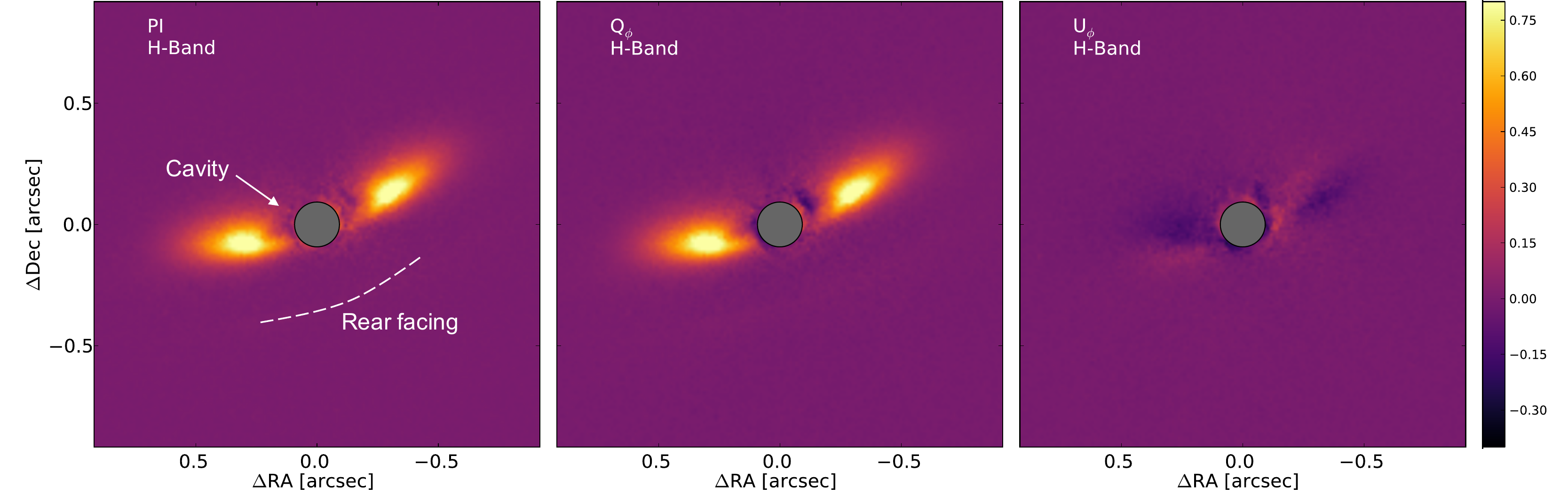}
\includegraphics[width =18cm]{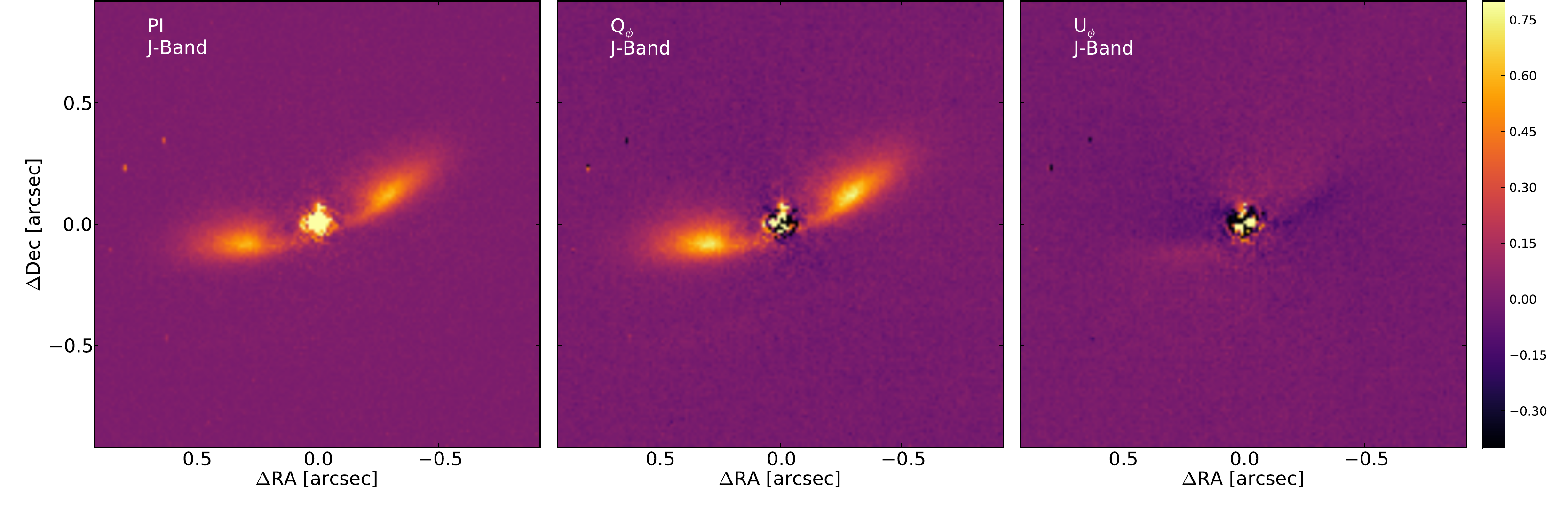}
\caption{Normalized polarized intensity map (left panel), Q$_\phi$ (middle), and U$_\phi$ (right) maps of J1608. Top: H-band observations, obtained with a coronagraph as illustrated by the gray circle of diameter 0.185\arcsec. The dashed line in the left panel traces the faint scattered light from the rear-facing near side of the disk (see Fig.~\ref{fig:J1608_faintline}). Bottom: J-band observations, obtained without coronagraph. Each Q$_\phi$ and U$_\phi$ maps are normalized to the maximum of Q$_\phi$.}
\label{fig:data_J1608}
\end{figure*}

\begin{figure*}
\centering
\includegraphics[width =18cm]{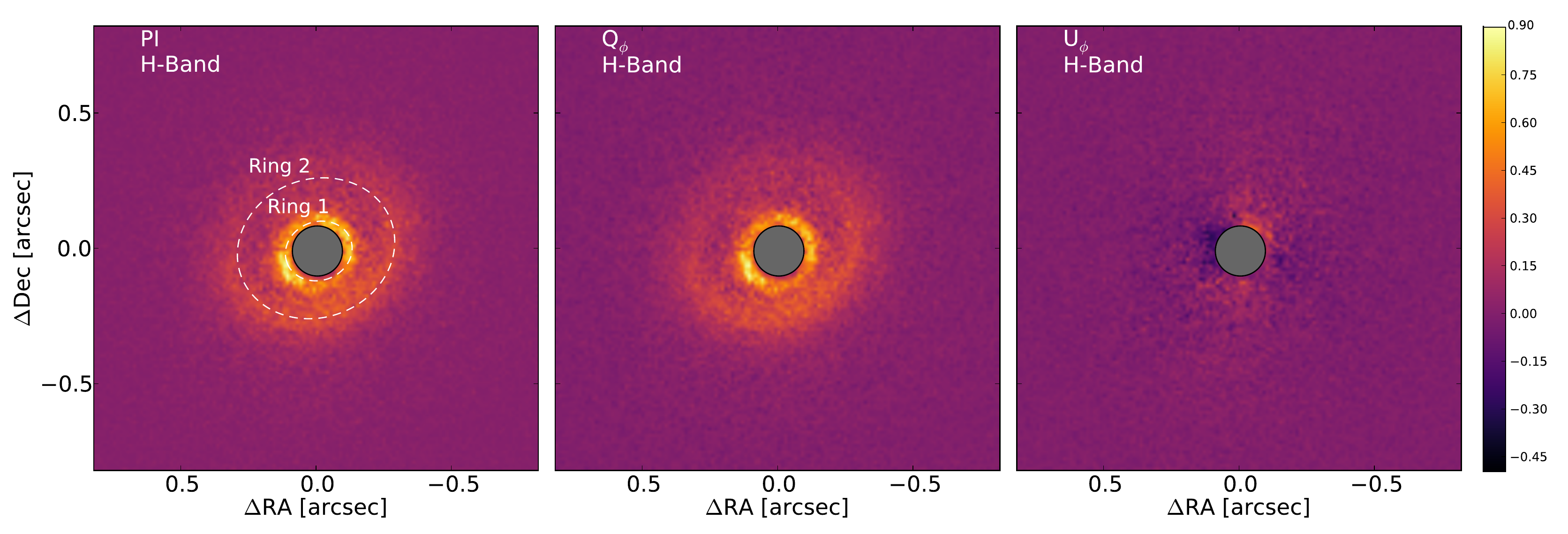}
\caption{Normalized polarized intensity map (left panel), Q$_\phi$ (middle), and U$_\phi$ (right) maps of J1852, observed in H-band with a coronagraph. The Q$_\phi$ and U$_\phi$ maps are normalized to the maximum of Q$_\phi$. The dashed ellipses in the left panel represent the two scattered light rings (see text for details).}
\label{fig:data_J1852}
\end{figure*}

\subsection{Results}
Our reduced images are shown in Fig.~\ref{fig:data_J1608} and Fig.~\ref{fig:data_J1852} for J1608 and J1852. 
J1608 appears to be very inclined, in images both with and without the coronagraph. We detect two lobes southeast and northwest of the star, as well as a faint line of scattered light\footnote{We show the faint southwest emission with a more favorable dynamic range in Fig.~\ref{fig:J1608_faintline}.} to the southwest that we interpret as the rear-facing side of the disk closest to us. The presence of this line will be discussed further in the modeling in Section 4.3. 
In the Q$_\phi$ maps, the disk shows emission above the noise level up to 0.54\arcsec{}~along the major axis. The central cavity is clearly visible in both the coronagraphic and non-coronagraphic images. We note that there is no clear emission from the northeast of the star, corresponding to emission from the far side of the disk being scattered backwards toward us.  In Section 4.3, we will focus on the modeling of the coronagraphic H-band image as the data were taken in better observing conditions, have a higher signal to noise ratio, and look very similar to the non-coronagraphic J-band data. 

In the coronagraphic images of J1852 (Fig.~\ref{fig:data_J1852}), we observe two rings in the $Q_\phi$ map, with peak values located at 0.125\arcsec{} and~0.295\arcsec{} from the position of the star along the major axis. The inner ring is cut off by the coronagraph and the peak radius may lie within this. The outer ring seems to be slightly off-centered compared to the inner ring ($\sim 0.01\arcsec{}$, 1\,pixel), likely an effect of the inclination and flaring of the disk surface~\citep[e.g.,][]{Boer_2016, Ginski_2016, Avenhaus_2018}. This is modeled in Section 4.4. The disk displays emission above the noise level as far as~0.4\arcsec{}along its major axis.

\subsection{Complementary data}

\paragraph{ALMA archival data.}

J1608 was observed as part of a large survey of disks in the Lupus clouds. In this work, we use the Band~6 observations (Project ID: 2015.1.00222.S, PI:\,Williams) obtained on July 23, 2016, at a resolution of~0.24\arcsec{}$\times$0.23\arcsec{}, cleaned with a Briggs robust weighting parameter of~+0.5. The data reduction of the continuum and line emission is presented in details in \citet{ansdell_2018}. 

\begin{figure}
    \centering
    \includegraphics[width = 8cm, trim={1.8cm 1cm 2cm 0cm}]{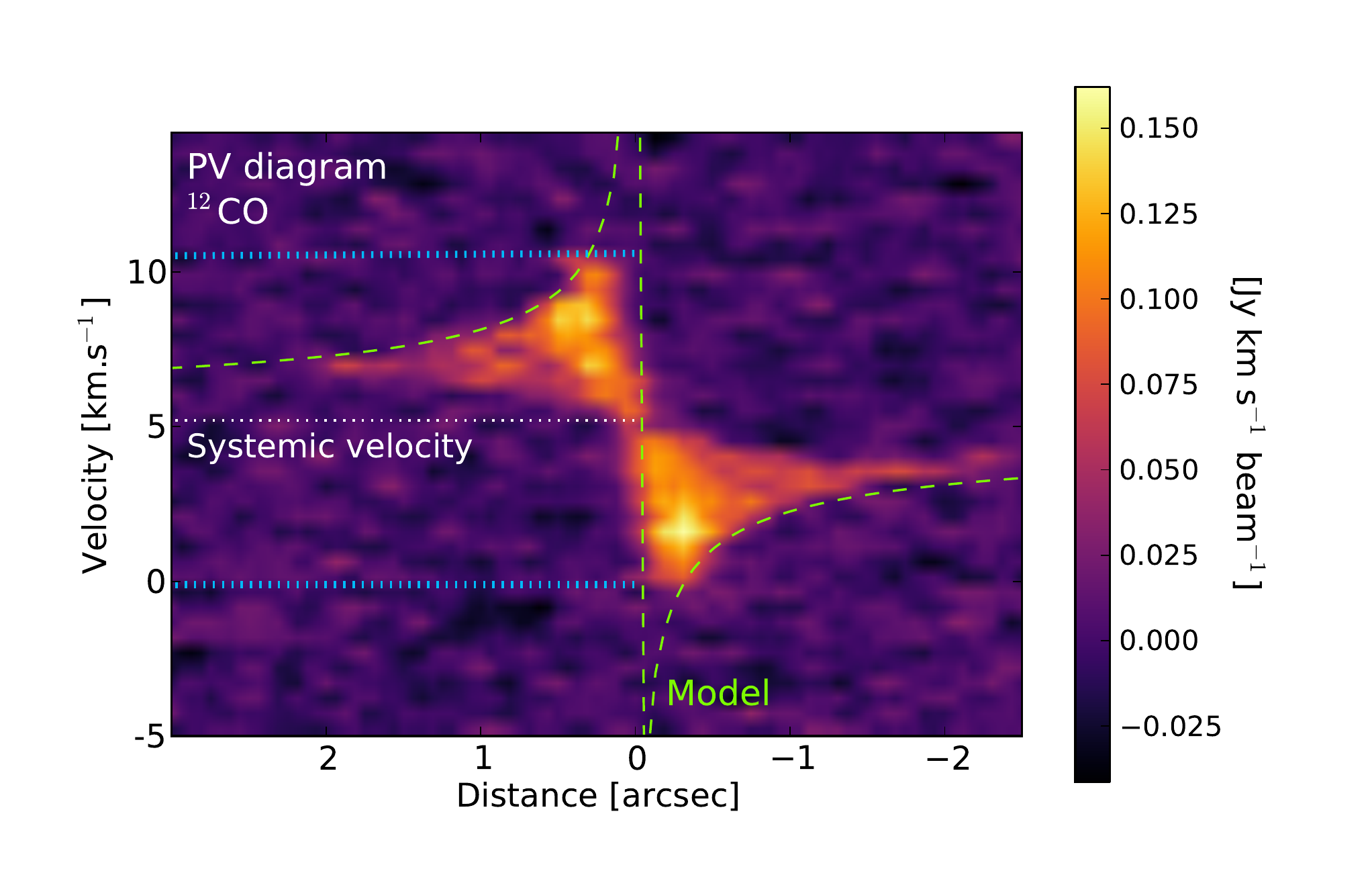}
    \caption{Position-velocity diagram of the  $^{12}$CO 2-1 emission in~J1608. The green dashed line corresponds to the Keplerian velocity of a disk at~74$^\circ$ of inclination around a 1.4 M$_\odot$ star. The highest velocity with emission is about 5.3 km\,s$^{-1}$ from the systemic velocity. It corresponds to a gas cavity radius of $\sim$48\,au, and is indicated by the blue lines.}
    \label{fig:diagramme_PV}
\end{figure}

We show the position-velocity (PV) diagram of the~$^{12}$CO 2-1 transition in Fig.~\ref{fig:diagramme_PV}, obtained with a velocity resolution of 0.11\,km\,s$^{-1}$. From the PV diagram, we retrieve the systemic velosity of the source to be~+5.2$\pm$0.4\,km\,s$^{-1}$~(LSR). We also note that no velocity higher than 5.3\,km\,s$^{-1}$ with respect to the star is detected, indicating the presence of an inner cavity in the gas. Modeling the disk velocities with Keplerian motion (assuming M$_\star$=1.4\,M$_\odot$, i=74$^\circ$), we infer that the inner radius of the gas cavity is $\sim$~48\,au, in  agreement with the outcome of thermo-chemical modeling of the CO observations~\citep[47\,au;][]{van_der_marel_new_2018}.

The ALMA continuum image is shown in the left panel of~Fig.~\ref{fig:prediction_J1608}. The large axis ratio indicates that the disk is highly inclined. Moreover, the presence of two blobs with higher intensity along the major axis, located at 0.4\arcsec{} from the star, denote emission coming from an optically thin ring.\\

For J1852, we use ALMA data observed on September~22, 2016, in Band 3 ($\sim$ 3\,mm, Project ID: 2015.1.01083.S, PI: Morino). The four continuum spectral windows were centered respectively on 91.5\,GHz, 93.4\,GHz, 101.5\,GHz and 103.5\,GHz. We used the~CASA pipeline to calibrate the data and extracted the continuum images using the CASA \texttt{clean} task, with a Briggs robust parameter of +0.5. The final image, after performing phase only self-calibration, is presented in the left panel of~Fig.~\ref{fig:prediction_J1852}, with the achieved beam of 0.38\arcsec$\times$0.31\arcsec. The image shows one unique ring, peaking at 0.3\arcsec~along the major axis. 

\paragraph{Spectral energy distributions.}
We compiled the SED of~J1608 using Vizier, to which we add the millimeter fluxes at 0.88\,mm and 1.3\,mm obtained from \citet{ansdell_2016} and \citet{ansdell_2018}, respectively. The SED of J1852 was computed using the VO SED Analyser\footnote{http://svo2.cab.inta-csic.es/theory/vosa50}~\citep[VOSA,][]{Bayo_2008}. We complement this SED with the low-resolution Spitzer/IRS spectrum from the CASSIS database \citep{Lebouteiller_2011}. The SEDs are shown on Figs.~\ref{fig:sed_J1608} and \ref{fig:sed_J1852}.

The SEDs of the two disks show a steep increase longwards of~20\,$\mu$m, typical of transition disks. Although~J1852 does not show any clear near-IR excess, a silicate feature at~10\,$\mu$m is present, characteristics of the presence of small hot grains close to the star \citep{Silverstone_2006,hughes_2010}.

\section{Radiative transfer modeling}
\label{sec:model}

\subsection{Methodology}
In order to understand the physical structure of the two~observed TDs, we construct models using the radiative transfer code \texttt{mcfost}~\citep{pinte_2006,pinte_2009}.  This code computes the thermal structure of the disk using a Monte Carlo method and produces images by ray-tracing. An iterative process is done to find a model that reproduces well the~SED and the images.  Considering the complexity of our observations, we do not aim to find the best-fit model, but rather a representative one. For each set of parameters, we compute the SED, the millimeter continuum image as well as the near-IR polarized intensity, Stokes Q and U maps. 
 All images are initially generated with infinite angular resolution, and then convolved with a point spread function reference image~(PSF; in practice, a non-coronagraphic or~FLUX image of the star) or the ALMA 2D Gaussian beam.

We also add noise to our scattered light predictions. We estimate it at each point of the map from the U$_\phi$ image of our observations \citep[as in][]{Muro-Arena_2018}. For each point in the U$_\phi$ map, we consider an aperture in which we determine the root mean square (rms) of the intensity. The aperture is taken as 4\,pixels of diameter, which is close to the FWHM of the point spread functions of each target (3.5\,pixels for J1608 and 4.3\,pixels for J1852). As our sources are inclined, the U$_\phi$ image might contain some physical signal~\citep{Canovas_2015}. Thus, in each aperture we subtract its mean value (physical signal) before extracting the rms. For each corresponding pixel in the model image, we simulate the noise by adding a Gaussian random number of the same rms centered on 0.

Since we will model emission very close to the coronagraph in the J1852 image, we apply a 2D attenuation map due to the coronagraph (Wilby et al. in prep). This numerical mask removes all signal inside the coronagraph mask radius (0.093\arcsec), with gradually decreasing attenuation down to 5\% at a radius of 0.15\arcsec. Finally, we compute the PI, Q$_\phi$ and U$_\phi$  maps using equations~(\ref{eq:PI}), (\ref{Qphi}), and (\ref{Uphi}). In this work, we compare the observed and synthetic PI images. A comparison between the observed and predicted Q and U as well as the Q$_\phi$ and U$_\phi$ maps are presented in~Appendix~\ref{sec:Qphi_Uphi}.

\subsection{Model setup}
We define various axisymmetric disk zones to reproduce the observed features of the two disks and assume a gas-to-dust ratio of 100.  For each region, we define the disk height as a power-law:
\begin{equation}
\label{eq:scale_height}
H(R) = H_\mathrm{100\,au} (R/100\,\text{au})^\beta
\end{equation}
where $\beta$ is the flaring exponent, R the radius and H$_\mathrm{100\,au}$ is the scale height at a radius of 100~au. A simple description of the surface density profile is adopted for each region of our disks, with a single power law: $\Sigma(R) \propto R^{p}$. In all our modeling we choose $p$\,=\,-1. We use astronomical silicates~\citep[similar to those shown in Fig. 3 of][]{Draine_lee_1984} with a number density described with a power law of the grain size $\mathrm{dn(a)} \propto a^{-3.5}\,\mathrm{da}$~\citep{Mathis_1977}. 
For each region, our free parameters are the inner and outer radius~(R$_\mathrm{in}$-R$_\mathrm{out}$), the dust mass and the scale height~(H$_\mathrm{100\,au}$). 

For both disks, we mimic dust settling by modeling separately the extents of small (0.01 - 0.5 $\mu$m) and large grains~(10 - 1000 $\mu$m). We fixed the flaring exponent $\beta$ to~1.1 for all type of grains.  The results are summarized in~Table~2 and described in detail in the following two subsections. 

\subsection{Modeling J1608}

\begin{figure*}
\includegraphics[width =18cm]{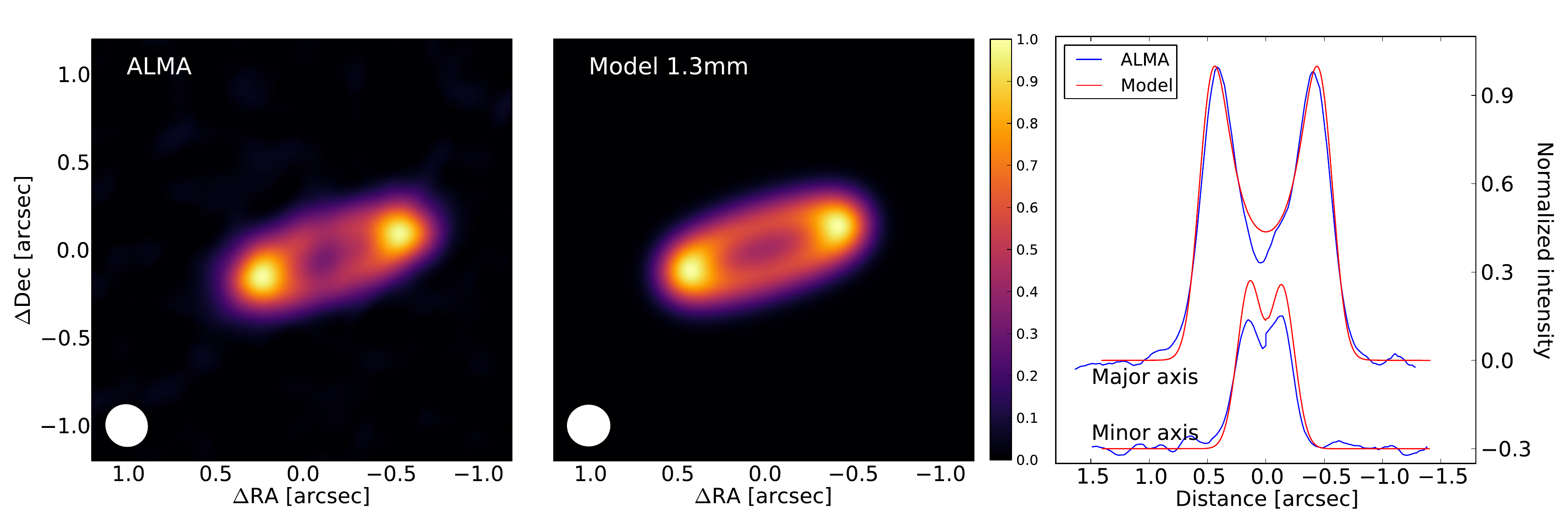}
\caption{Left: Normalized ALMA Band 6 image of J1608, with a peak flux of 7.1\,mJy. The beam is shown in the bottom left of the image. Middle: Normalized model obtained after convolution by a beam of 0.24\arcsec{}$\times$0.23\arcsec{}. Right: Radial cuts along the major and minor axis, with each map normalized to its maximum. For clarity, the cut along the minor axis is shifted by -0.3 in normalized intensity. The peaks are located at 0.4\arcsec (62\,au) along the major axis and~0.14\arcsec (22\,au) along the minor axis.}
\label{fig:prediction_J1608}
\end{figure*}

\begin{figure*}
\centering
\includegraphics[width =18cm]{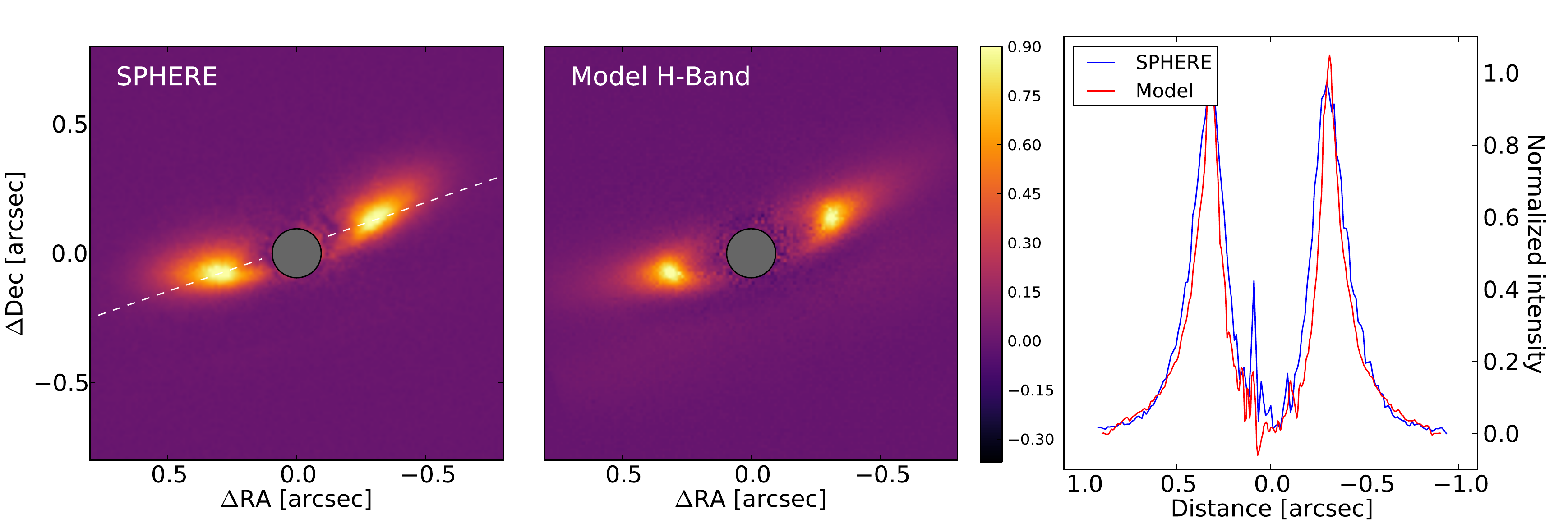}
\caption{Left: Observed polarized intensity of J1608. Middle: PI model map. Both maps are normalized to their maximum. Right: Radial cuts along the major axis for the PI images of the data and model, smoothed with a Gaussian kernel with a standard deviation of 9 mas. The position of the cut is indicated by the dotted line in the left panel.}
\label{fig:model_J1608}
\end{figure*}
 
\paragraph{Large grains.}
We first aim to reproduce the thermal emission detected in the ALMA Band 6 data with a ring of large grains, adjusting the dust mass to fit the total millimeter flux. Our convolved model prediction at 1.3\,mm is presented in the center panel of Fig.~\ref{fig:prediction_J1608}. The right panel of this figure shows radial cuts along the major and minor axis. In our model, the radial position of the maxima as well as the radial extent in each cut are well reproduced. However, the model overestimates the flux in the inner region of the disk~(i.e., inside the gap) by about 30\%.

To reproduce the observed aspect ratio of the disk, a high inclination ($\sim74^\circ$) is needed. However, we find that the inclination should not be larger than 80$^\circ$, otherwise the photosphere would be occulted at short wavelengths, in contradiction with the shape of the SED~\citep[see for example the SED of the edge-on disk ESO-H{\ensuremath{\alpha}}\,569;][]{Wolff_2017}.
The radial width of the ring made of large grains has to be sufficiently small that the position and shape of the maxima of the cut along the major axis match the data. We find that a radial width for the large grains ring of about 10\,au is consistent with the data, with~R$_\mathrm{in}\sim$~77\,au. However, a narrower ring would still reproduce the observations.

Finally, the fluxes of the peaks and depth of the gap along the minor axis  depend both on the radial width and the vertical thickness of the disk. 
If the scale height of the large grains is too large, after convolution by the beam, the two sides of the ring would appear as connected, leading to a flat intensity profile along the minor axis. On the other hand, if the zone of large grains is too thin vertically for a given disk mass, the fluxes at the peaks of the minor axis cut would become too large.  The appropriate scale height in our modeling is between 3 and 5\,au at 100\,au, which is similar to the value obtained for HL\,Tau~\citep[1\,au at 100\,au,][]{Pinte_2016}. We note that a smaller scale height, such as in HL\,Tau, can not be excluded by our model. We show the non-convolved model in Appendix~\ref{sec:no_convolution}. All the model's parameters are presented in Table~\ref{tab:parameters}.

\begin{figure}
 \resizebox{\hsize}{!}{\includegraphics{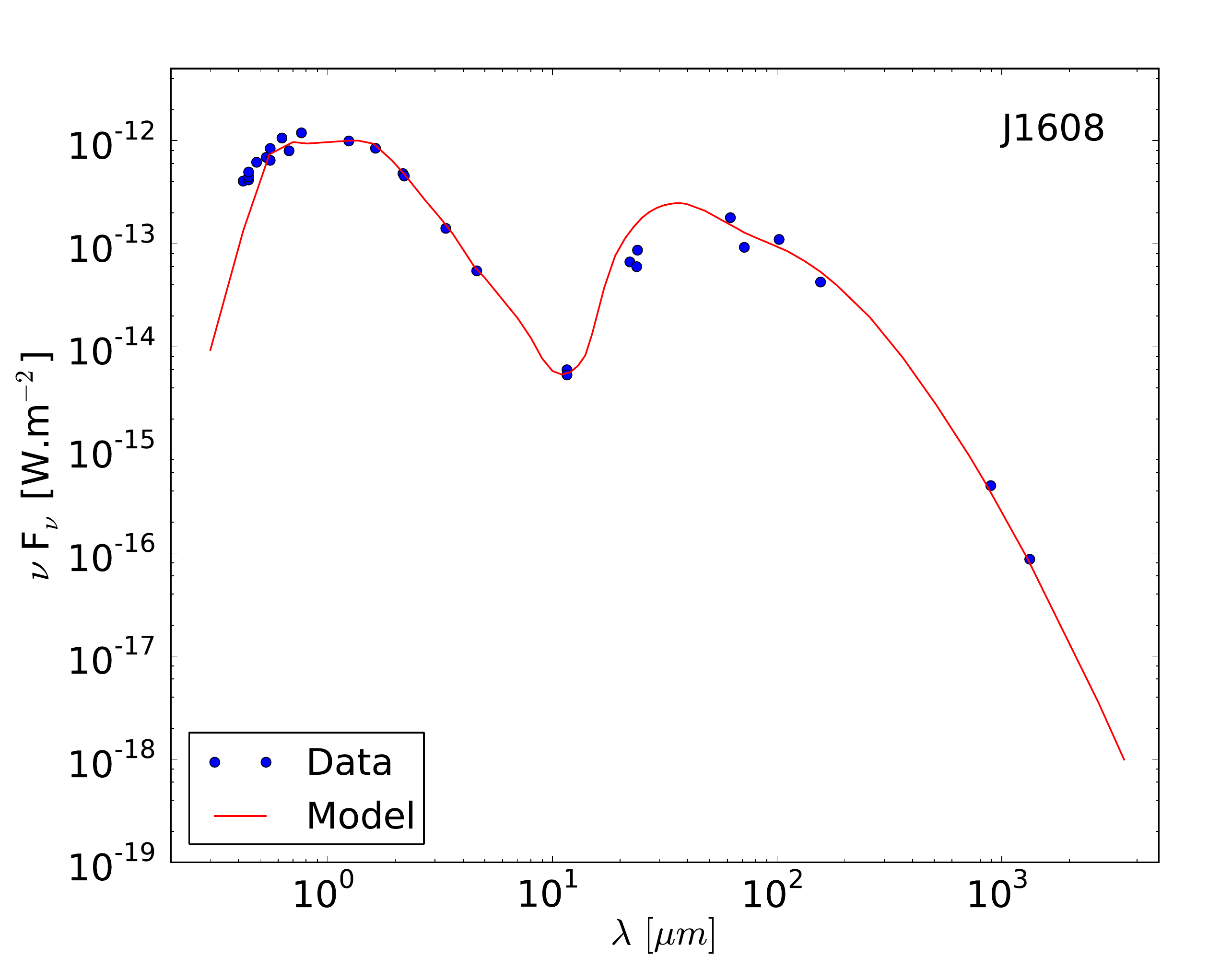}}
\caption{Spectral energy distribution of J1608 (blue circles) and our model prediction (red line). The model is corrected by the~Av.}
\label{fig:sed_J1608}
\end{figure}

\paragraph{Small grains.}
The synthetic scattered light image computed assuming the same spatial distribution for small grains and large grains does not provide a good match to the images. To reproduce the observations, the radial extent of small grains and their scale height need to be larger than those of the large grains, from 50 to 150\,au radially, and about~12\,au vertically at 100\,au.  We note that this  layer is already included in the millimeter predictions displayed in~Fig.~\ref{fig:prediction_J1608}, but does not contribute significantly at millimeter wavelengths. The scattered light image of our model is presented in~Fig.~\ref{fig:model_J1608}, with a cut along the major axis direction. The cuts were smoothed with a Gaussian kernel with a FWHM of~50\% the measured image resolution~($\sim$ 10\,mas).
The inner radius of the small grains distribution is very close to that of the gas, as estimated from the PV diagram (48\,au, see Fig.~\ref{fig:diagramme_PV}), consistent with the expectation that these grains are well coupled to the gas.

We note that in the observation, no scattered light is detected from the northeast part of the disk (Fig.~\ref{fig:model_J1608}).
Geometrically, the existence of the bottom line located in the~southwest suggests that it traces the part of the disk nearest to us, while the northeast region would correspond to the more distant side of the disk. 
This implies that the phase function of the polarized intensity is such that there is no or very little backward scattering (on the assumption that the disk is axisymmetric).  
The polarized intensity is the product of the polarization degree and the total intensity. While the polarization degree from backward scattering is similarly low for all grain size \citep[see Figure~7 of ][]{Keppler_2018}, the phase function of the intensity varies significantly with the grain size.
We reproduce this feature by selecting grains smaller than 0.5\,$\mu$m. Grains larger than~1.2\,$\mu$m also have a very small efficiency for backward scattering. However, with such grains our models showed that the position of the maximum intensity along the major axis is independent of the inner radius position and located further out than seen in the data.  These grains are therefore not compatible with the observations. For grains with an intermediate size~(between 0.5 and 1.2\,$\mu$m), backward scattering is very efficient, which would lead to a significant signal northeast of the disk coming from its far side.

We show the SED compared to that of the model in~Fig.~\ref{fig:sed_J1608}. All the model parameters are presented in Table~\ref{tab:parameters}, and a schematic representation of our model is shown in the top panel of~Fig.~\ref{fig:scale_heights}. 

We note that we are able to recreate the same general structure as seen in J1608, with two lobes along the major axis, no emission from the northeast region of the disk~(backward scattering) and a faint line south of the disk, but these characteristics are not perfectly reproduced. For example the southern line is brighter in our model than in the observations (see Fig~\ref{fig:J1608_faintline}).  To reproduce better this part of the image, the layer of small grains scattering light must be thinner vertically, but this in turn has an impact on the extent of the lobes.  Moreover, the southern line is closer to the star than in the data. A way to push this line further would be for example to increase the mass of the scattering grains, the scale height or the inclination. However, each case would lead to an extinction of the photosphere that is not observed in the~SED. 
Thus the model that we present here is a compromise to reproduce both the ALMA and SPHERE images, together with the SED. This should be considered as a working model to derive the main structural characteristics of the J1608 system. These are: 1/~a high inclination, 2/ large grains more concentrated vertically and radially than the small grains, and 3/ a distribution of size in small grains that produces low polarized intensity in backward scattering.

\begin{table}
\centering
\caption{Parameters for our radiative transfer models.}
\begin{tabular}{|rl|c||cc|}
\hline
& &J1608 & \multicolumn{2}{|c|}{J1852}\\
\hline
Inclination&[$^\circ$]&74&\multicolumn{2}{|c|}{30}\\
PA&[$^\circ$]& 19 & \multicolumn{2}{|c|}{34}\\
\hline
&&&\multicolumn{2}{c|}{Inner disk}\\
$a_\mathrm{min}$-$a_\mathrm{max}$& [$\mu$m] &-&\multicolumn{2}{|c|}{$0.01-5$}\\
$R_\mathrm{in}$-$R_\mathrm{out}$ &[au]&-&\multicolumn{2}{|c|}{$0.1-5$}\\
Mass &[M$_\odot$]& -&\multicolumn{2}{|c|}{$9\cdot10^{-11}$}\\
H$_\mathrm{100au}$ &[au] & -&\multicolumn{2}{|c|}{0.9}\\
\hline
&&Small grains&\multicolumn{2}{c|}{Small grains}\\
$a_\mathrm{min}$-$a_\mathrm{max}$& [$\mu$m]&$0.01 - 0.5$&\multicolumn{2}{c|}{$0.01-0.5$}\\
$R_\mathrm{in}$-$R_\mathrm{out}$ &[au]&$50-150$&$15-22$&$42-65$ \\
Mass  &[M$_\odot$]& $2\cdot10^{-6}$&$2\cdot10^{-8}$&$7\cdot10^{-7}$\\
H$_\mathrm{100au}$ &[au]  & 12&\multicolumn{2}{c|}{15}\\
\hline
&&Large grains&\multicolumn{2}{c|}{Large grains}\\
$a_\mathrm{min}$-$a_\mathrm{max}$& [$\mu$m] &$10 - 1000$&\multicolumn{2}{c|}{$10-1000$}\\
$R_\mathrm{in}$-$R_\mathrm{out}$ &[au]&$77 - 87$&$15-22$ & $42-65$\\
Mass&  [M$_\odot$]& $5\cdot10^{-5}$& $3\cdot10^{-7}$ & $7\cdot10^{-5}$\\
H$_\mathrm{100au}$ &[au]  & 5& \multicolumn{2}{c|}{1}\\
\hline
\end{tabular}
 \tablefoot{Each parameter was adjusted during the modeling, except for the grain size ($a_\mathrm{min}$-$a_\mathrm{max}$).}
\label{tab:parameters}
\end{table}

\begin{figure}
    \centering
 \resizebox{\hsize}{!}{\includegraphics{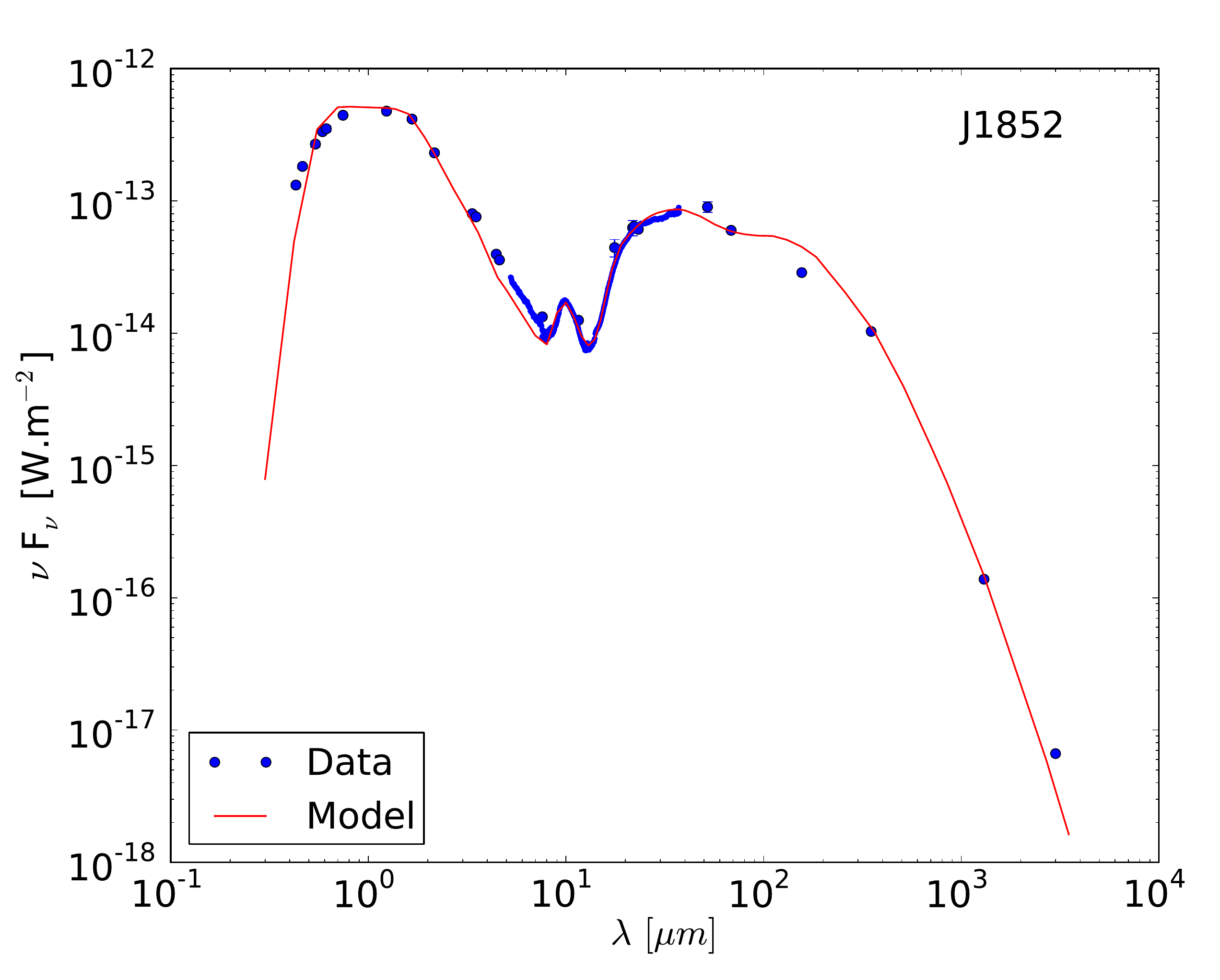}}
\caption{Spectral energy distribution of J1852 (blue circles) and our model (red line). The model is corrected by the Av.}
\label{fig:sed_J1852}
\end{figure}

\begin{figure*}
\includegraphics[width =18cm]{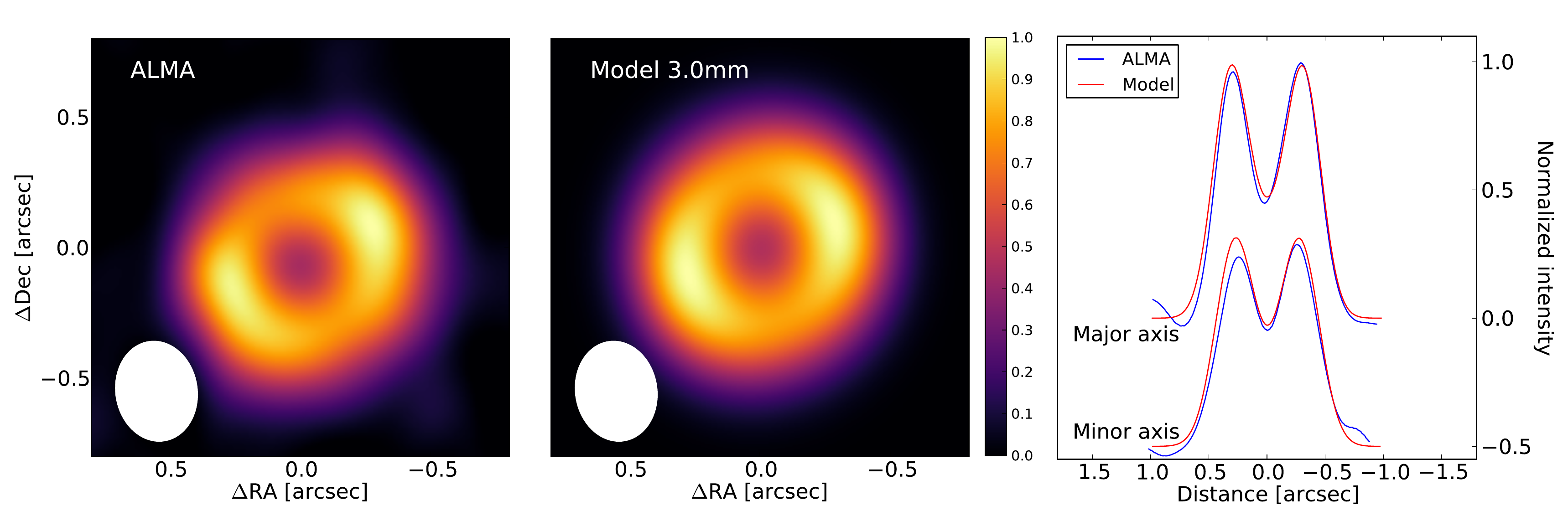}
\caption{Left: Normalized ALMA Band 3 image of J1852 (peak flux is 1.5\,mJy). The beam is represented in the bottom left of the image. Middle: Normalized model obtained after convolution by a beam of 0.38\arcsec{}$\times$0.31\arcsec{}. Right: Cuts along the major and minor axis, with each map normalized to its maximum. For clarity, the cut along the minor axis is shifted by -0.5 in normalized intensity. The peaks are located at 0.3\arcsec (44\,au) along the major axis.}
\label{fig:prediction_J1852}
\end{figure*}

\begin{figure*}
    \centering
    \includegraphics[width =18cm]{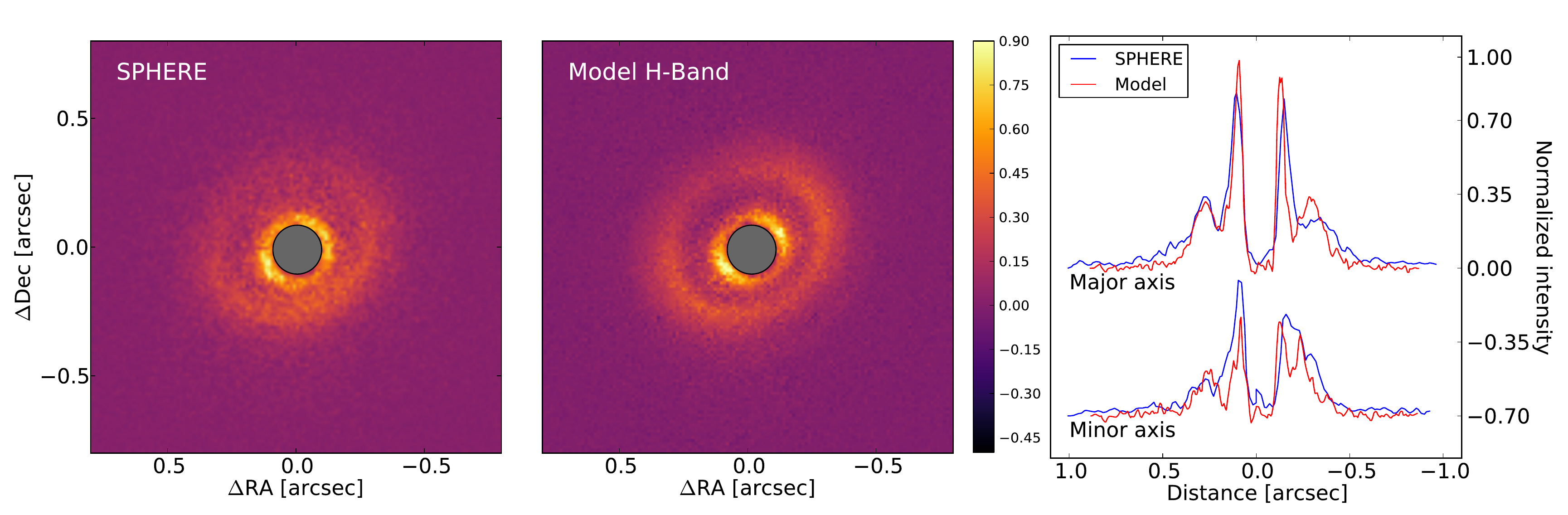}
    \caption{Left: Observed PI image of J1852. Middle: PI model map. Both maps are normalized to their maximum. Right: Radial cuts along the major and minor axis for the PI images of the data and the model. Each curve is convolved with a Gaussian kernel with a standard deviation of 11 mas.}
    \label{fig:model_J1852}
\end{figure*}

\subsection{Modeling J1852}

The SED of J1852 shows a steep increase around~20\,$\mu$m, typical of transition disks, due to the outer edge of the cavity. A clear silicate feature at 10\,$\mu$m is present, characteristic of small hot grains close to the star.  While scattered light image shows signal just outside of the coronagraph, there is no corresponding clear millimeter emission at the same location. The scattered light image shows a second bright ring at 0.295\arcsec, which does correspond to the ring detected in the millimeter image. 

To reproduce the scattered light and SED features, we consider three zones in our model: 1/~a tenuous inner disk region, solely required to reproduce the silicate feature, 2/~an inner ring, responsible for the sharp jump in the mid-IR SED and for the emission seen in the SPHERE image just outside of the coronagraph radius, and finally, 3/ an outer ring, to account for the second brightness increase in the near-IR polarized image and the millimeter ring. The structure is illustrated in the bottom panel of Fig.~\ref{fig:scale_heights}. The three zones are described in the following sections.

\paragraph{Inner disk.}
The inner region, responsible for the silicate emission around 10\,$\mu$m in the SED, is modeled by small grains~(0.01-5\,$\mu$m) between 0.1 and 5\,au, although this is not well-constrained. To reproduce the shape of the emission, we consider a mixture of silicate, composed at 65\% of olivine~\citep{Dorschner_1995} and~35\% of astronomical silicates~\citep{Draine_lee_1984}.

\paragraph{Inner ring.}
The position of the wall in the SED traces the dust temperature, which is dependent on both the inner radius and the size of the dust. For simplicity, we chose to use the same grain size distribution as for J1608. As the inner radius of J1852 is likely located behind the coronagraph, we constrained it using the SED. We reproduce well the inner ring by defining its radial extent between 15 and~22\,au, with a scale height of 15\,au at 100\,au. 

\paragraph{Outer ring.}
To reproduce the peak of intensity around~0.295\arcsec{}~($\sim$43\,au), an increase of the surface density in the small grains is needed, between 42 and 65\,au. 
The mass of this region is adjusted to reproduce the relative brightness of the rings and the SED. The scale height is fixed to be the same as that of the inner ring.

Our model reproduces relatively well the position and the brightness of the peaks in the major and minor axis directions. The second ring shows an offset to the center as seen in the data. However, we were not able to reproduce the surface brightness inside the gap between the inner and outer ring in the southern part of the disk. This region is twice brighter in the data than in the model. In our model the gap is empty and the ring edges are sharp, while in practice, a low surface density of small grains could be filling it. 
 
 \paragraph{Large grains.}
Unlike J1608, we do not have sufficient high resolution data to show that the large grains are radially distributed differently to the small grains, so for now we assume that they are radially co-located. We adjusted the total mass of the large grains to reproduce the observed SMA flux at~1\,mm and the ALMA 3\,mm flux in the~SED, allocating a small fraction of mass to the inner ring to account for the scattered light data. The large beam considered in this work also dilutes the emission of the inner ring~(as modeled here), making it very difficult to detect~(see Appendix~\ref{sec:no_convolution}). 
 With our assumptions on the radial structure and grain composition of the large grain population, we inferred its scale height from the effect on the 100\,$\mu$m emission. Indeed when the scale height is larger, as more grains receive light from the star, they warm up and emit more at~100\,$\mu$m~\citep[][]{Dullemond_2004, Woitke_2016}.  To fit the SED, it was necessary to reduce the scale height of the large grains to about~1\,au at 100\,au. However, it should be noted that modeling the SED is degenerate. In particular, changing the minimum size of the large grain population~(a$_\mathrm{min}$ in Table~\ref{tab:parameters}) from 10 to 300\,$\mu$m, we find that the small and large grain populations could be distributed similarly and share the same scale height, while leading to similar excess and images as in the previous model.  
 
\section{Discussion}
\label{sec:discussion}
\subsection{Dust vertical settling}
Our radiative transfer modeling of J1608 indicates that small and large grains have a different spatial distribution~(see~Table~\ref{tab:parameters}).  Small grains~($<1\,\mu$m) are found to be more extended vertically than large grains~($>10\,\mu$m). 
Small grains are found up to a height of 15\,au at~100\,au, in agreement with other results around~T\,Tauri stars~\citep[e.g.,][]{Burrows_1996, Wolff_2017, Pohl_2017}.
In contrast, to reproduce the ALMA image and SEDs, the height of the large grain layer at 100\,au has to be of a few astronomical units.

The need for stratification of dust grains was already suggested in earlier studies. \citet{Duchene_2003} showed that a perfect mixing of dust grains is not able to reproduce simultaneously HST scattered light images and IRAM millimeter images of HK\,Tau\,B, implying that vertical settling is occurring. Such a radial stratification of the dust distribution was also highlighted by \citet{Pinte_2007} while modeling IR and optical scattered light images of GG\,Tau.  In all cases, small grains are inferred to be located at the disk surface while large grains are found closer to the mid-plane, a natural outcome of vertical settling. 

Vertical settling of particles occurs simultaneously with radial drift due to the effect of stellar gravity and gas drag on the dust \citep[][]{Weidenschilling_1977}. Because the pressure force acts only on the gas, the gas  rotates at sub-Keplerian speed, while decoupled dust grains rotate faster, at Keplerian speed. The gas drag on the dust particles leads to inward drift.
Moreover, through the stellar gravity and the interaction with the surrounding gas, dust grains settle onto the midplane, with a different efficiency depending on the coupling of the grains.
Small grains are well coupled to the gas and hence are located at similar scale heights; large grains are relatively decoupled from the gas, and settle  to the midplane \citep{Barriere-fouchet_2005}. \citet{Laibe_2014} found that vertical settling is much faster than the radial drift of the particles, even when taking into account grain growth.

\citet{Fromang_2006} carried out ideal magneto-hydrodynamic (MHD) simulations to quantify the effect of magnetic field on dust settling. They consider a  strongly magnetized and turbulent disk, with a viscous coefficient of $\alpha\approx 1.5\times10^{-2}$ in the simulation, and no grain growth.
They find that the scale height of 10\,cm bodies is about $H_\mathrm{100\,mm} \approx 0.23\, H_\mathrm{gas}$~\citep[equation 43 of][]{Fromang_2006}, which seems too large to account for current estimates from observations. If we assume that scattered light traces the gas and that all light at~1.3\,mm is emitted by grains of similar size \citep{Draine_2006}, the modeling of J1608 gives mm dust scale heights on the order of~$H_\mathrm{1.3\,mm}=0.41\,H_\mathrm{gas}$. This is similar or smaller than the predictions of \citet{Fromang_2006} for particles~100 times larger, which are expected to be considerably more settled. 

On the other hand, \citet{Dullemond_2004} showed analytically that settling is more efficient for disks with low turbulence. 
It is also is more efficient if the grain size distribution contains fewer small grains or if the gas-to-dust ratio is low~\citep{Mulders_Dominik_2012}. 
Moreover, in the outer parts of the disk the ionization fraction might be such that a perfect coupling to magnetic field is unlikely, and non-ideal effects might be expected.  
In this context, \citet{Riols_2018} showed that ambipolar diffusion allows much more efficient settling of large grains than in perfect MHD models, allowing to reproduce the constraints on~HL\,Tau. 

\subsection{Dust radial distribution}
\paragraph{Outer radius.}
Our model of J1608 shows that small grains extend to larger radii than large grains (Table~\ref{tab:parameters}).
This is expected as small grains are predicted to be well coupled to the gas, well detected beyond the ring seen in the ALMA image. On the other hand, large grains are (partially) decoupled from the gas, and experience radial drift and dust trapping. Earlier studies have shown that the gas outer radii extend further than the ones measured in the millimeter continuum. An average ratio between the gas and dust outer disk radii of 1.96$\pm$0.04 was found in~22 disks of the Lupus star forming region \citep{ansdell_2018}. This trend is also observed in 12 disks in Chamaeleon\,II~(Villenave et al. in prep) and in individual disks such as PDS\,70 where the gas is detected up to 160\,au and millimeter dust up to~110\,au~\citep{Long_2018}. 
 We note, however, that the sensitivity limits between the ALMA observations in CO and those in polarized scattered light are different. The outermost radius at which scattered light is detected depends on the stellar illumination (which drops as~r$^{-2}$).  Besides the sensitivity limits, the difference between gas/small grains and large grains can be due to optical depth effects combined with the radial drift of the large grains \citep{facchini2017}. 
Depending on the inclination of the system, the vertical height of small grains can also have a large effect on their detectable radius. If the vertical height of small grains decreases after some radius, the surface layers could be located in the shadow of the inner region and would not be detected in scattered light \citep{Muro-Arena_2018}. 

\paragraph{Inner radius.}
From the PV diagram shown in Fig.~\ref{fig:diagramme_PV} and our modeling, we found that both the small grains and gas in J1608 extend inwards of the large grains, indicating that the cavity is not completely empty. 
Likewise, an inner ring in scattered light is detected inside the millimeter cavity of~J1852. A similar conclusion was reached on several other transition disks, which show a CO cavity smaller than the millimeter dust cavity, such as Sz\,91~\citep{Canovas_2015, Canovas_2016} or RXJ1604.3-2130A~\citep[][]{Zhang_2014} for example. In each of these systems, the authors found the CO to extend at least 20\,au inward of the outer edge of the millimeter dust cavity.
These differences in inner radius could be related to the mechanisms responsible for the cavities in transition disks, which we explore in the next subsection. 

\subsection{Comparison with other transition disks}

Several mechanisms have been proposed to explain the origin of cavities in transition disks: photoevaporation \citep{Owen_2011}, dead zones \citep{Flock_2015, Pinilla_2016}, opacity effect via grain growth \citep{Dullemond_2005, Birnstiel_2012}, or planetary or stellar companion interacting with the disk~\citep{Crida_2007, Facchini_2013}. Recent studies showed that most of the transition disks studied at high angular resolution~(a sample biased toward the brightest objects) still have moderate accretion rates~\citep{manara_2014} and rather small~CO cavities. Thus, photoevaporation might not be the main mechanism in this sample~\citep{Pinilla_2018, van_der_marel_new_2018}.

Dead-zones are low ionization regions in which magneto-rotational instability is suppressed \citep{Blaes_Balbus_1994, Flock_2012}. In such regions the rate of gas flow decreases, leading to accumulation of gas at the outer edge of the region. The pressure bump is able to trap particles and stop the radial drift of large grains, which leads to a ring-like morphology for the millimeter dust as observed in transition disks \citep{Pinilla_2016}. 
Synthetic scattered light and millimeter continuum images of disk with a dead zone in the inner 30\,au show that the inner edge of the ring is located at about the same radius in both tracers. 
If a dead zone and a MHD wind act together however, a larger difference in inner radii could be observed.

Dust depleted cavities can also be generated by planets. Planets of mass larger than 1\,M$_\mathrm{J}$ can carve gaps in the gas and induce large perturbations in the gas surface density~\citep{Dong_fung_2017}, in turn generating pressure maxima that trap dust particles. In this case, the inner region of the disk is depleted in millimeter grains, while smaller grains can flow inside the planet's orbit, and potentially be detected in scattered light~\citep{Ovelar_2013,Dong_2015}. 

\begin{figure}
    \centering
     \resizebox{\hsize}{!}{\includegraphics{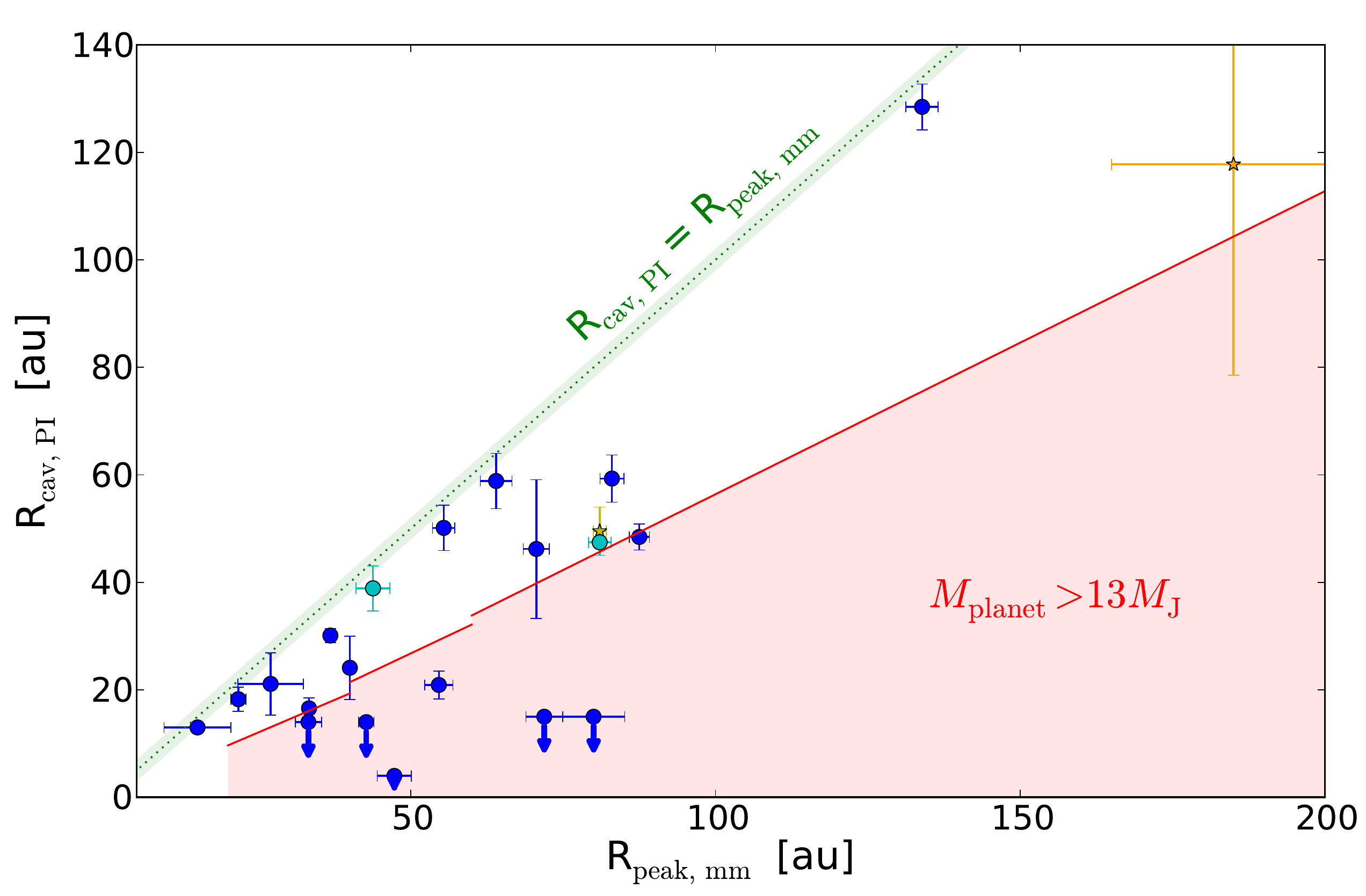}}
    \caption{Comparison of the cavity sizes as measured in the millimeter and in scattered light for a sample of 22 transition disks. The cyan points correspond to the two systems modeled in this paper, while the yellow star symbols refer to HD\,142527 that has a stellar companion in the cavity, and PDS\,70 where a planet has been detected.
    The upper limits, indicated with arrows, refer to systems for which scattered light cavities could not be measured down to the coronagraphic radius. Horizontal bars correspond to one tenth of the millimeter beam, while vertical bars represent the interval between the cavity radius and the peak in scattered light (Table~\ref{tab:comparison}). The red region shows the ratio for which the planet mass inferred with the prescription of \citet{Ovelar_2013} is larger than 13 Jupiter masses. The three breaks correspond to their models with planets at radii of 20, 40 and 60\,au, respectively.}
    \label{fig:comparison_sizes}
\end{figure}

\begin{table*}[]
    \caption{Position of millimeter and scattered light cavities for 22 transition disks, ordered by increasing millimeter cavity size.}
    \centering
    \begin{tabular}{|l|c|l||c|rlc|r|}
    \hline
    Source & d [pc]& Band& R$_\mathrm{peak,mm}$ [au]&&R$_\mathrm{in,PI}$  [au]&R$_\mathrm{peak,PI}$ [au]&Ratio\\
    \hline
    HD\,100546 & 110 $\pm$ 1 & B7 - R' &15.0  & &12.0& 14.0 & 0.81, 0.96\\ 
    HD\,169142 &114 $\pm$ 1& B6 - J& 21.7& &16.0&20.5& 0.74, 0.94\\
    \emph{V4046\,Sgr}& 72 $\pm$ 1&B6 - H&27.0&&15.3&26.9&0.57, 0.99\\
    UX\,Tau\,A & 140 $\pm$ 2 & B6 - H & 33.2 &<&14 && < 0.42\\ 
    HD\,100453 &104 $\pm$ 1&B6 - R'& 33.3 && 14.6&18.5&0.44, 0.55\\
    T\,Cha& 110 $\pm$ 1& B3 - H& 36.8 && 28.8&31.4& 0.78, 0.85\\
    HD\,143006&166 $\pm$ 4 & B6 - J&40.0&&18.2&30.0 & 0.45, 0.75 \\
    DoAr44& 146 $\pm$ 1 & B7 - H &42.7 & <&14 && < 0.33\\
    CQ\,Tau & 163 $\pm$ 2 & B6 - J& 47.3& <&4& & < 0.08\\
    J1852 & 146 $\pm$ 1& B3 - H& 43.8 & &34.7 &43.1& 0.79, 0.98\\
    HD\,135344B & 136 $\pm$ 2 & B7 - R & 54.6 && 18.3 & 23.5 & 0.34, 0.43\\
    HD\,97048& 185 $\pm$ 1 & B7 - J & 55.4 & &45.9 &54.4& 0.83, 0.98\\
    LkCa\,15& 159 $\pm$ 1&B7 - J & 64.0 & & 53.7&64& 0.84, 1.0\\ 
    Oph\,IRS\,48 & 134 $\pm$ 2 & B9 - Q & 70.6 & &33.3& 59.1 & 0.47, 0.84\\ 
    RY\,Lup & 159 $\pm$ 2 & B7 - H &  71.9 & <&15& &< 0.20\\
    MWC758 & 160 $\pm$ 2 & B7 - Y&80.0 & < &15&&< 0.19\\
    PDS\,70 & 113 $\pm$ 1 & B7 - J & 81.0 & & 45.0& 54.0 & 0.56, 0.67\\
    J1608 & 156 $\pm$ 6& B6 - H& 81.0 & & 45.0& 49.9  & 0.56, 0.82\\
    RXJ1604.3-2130A& 150 $\pm$ 1 & B6 - R' & 83.0 & &54.9& 63.7& 0.66, 0.77\\
    Sz\,91 & 159 $\pm$ 2 & B7 - Ks&87.5&&46.0& 50.9& 0.52, 0.58\\
    HD\,34282 & 312 $\pm$ 5 & B7 - J & 133.9 & &124.2 &132.7& 0.93, 0.99\\
    \emph{HD\,142527}& 157 $\pm$ 1&B7 - Ks & 165, 205& &78.5 & 157.0 & 0.38, 0.95\\
    \hline
    \end{tabular}
    \tablefoot{Known binary systems are indicated with the name in italic.  We report the peak of the millimeter intensity beyond the cavity, along with the position of the inner radius and peak in scattered light. When the cavity is not detected down to the coronagraph radius in scattered light, we use the symbol <. The position of the scattered light cavity (R$_{\rm{cav,PI}}$),  defined as the mean between R$_{\rm{in,PI}}$ and R$_{\rm{peak,PI}}$, is used in Fig.~\ref{fig:comparison_sizes}. 
    }
    \tablebib{HD\,100546: \citet{Pinilla_2018, garufi_2016}, HD\,169142: \citet{Fedele_2017, Pohl_2017b, Bertrang_2018}, V4046\,Sgr: \citet{Rosenfeld_2013, Avenhaus_2018}, HD\,100453: \citet{van_der_plas_2019, Benisty_2017}, T\,Cha: \citet{Hendler_2018,Pohl_2017}, UX\,Tau: \citet{Pinilla_2018}, Menard (in prep); HD\,143006: \citet{Benisty_2018, Perez_2018}, DoAr44: \citet{Casassus_2018, Avenhaus_2018}, CQ\,Tau: \citet{Pinilla_2018}; Benisty (in prep), J1852: This work,   HD\,97048: \citet{van_der_plas_2017a, Ginski_2016}, LkCa\,15: \citet{Andrews_2011, Thalmann_2015, Thalmann_2016},  HD\,135344B: \citet{Pinilla_2018, Stolker_2016}, Oph\,IRS\,48: \citet{van_der_Marel_2013, Bruderer_2014, Geers_2007}, RY\,Lup: \citet{Pinilla_2018, Langlois_2018}, MWC758: \citet{Marino_2015, Benisty_2015}, RXJ1604.3-2130A: \citet{Pinilla_2018b,Pinilla_2015},  PDS\,70: \citet{Long_2018,Keppler_2018}, J1608:\citet{ansdell_2016}, This work, Sz\,91: \citet{Canovas_2016, Tsukagoshi_2014}, HD\,34282: \citet{van_der_plas_2017b}, de Boer (in prep.), HD\,142527: \citet{Boehler_2017, Avenhaus_2014}. }
    \label{tab:comparison}
\end{table*}

To assess the origin of the cavity in transition disks, we compiled a sample of 22 disks that have both scattered light and millimeter observations, as presented in~Table~\ref{tab:comparison}. 
This sample includes two known binary systems,~HD\,142527 and~V4046\,Sgr \citep{Biller_2014, Quast_2000}. The excentric binary companion in HD\,142527 is likely responsible for the cavity \citep{Price_2018}, but this is probably not the case for the very close binary system V4046\,Sgr~\citep[2.4\,days period,][]{Quast_2000,dorazi2018}.
All sources show a resolved dust cavity in the millimeter and, except Oph\,IRS\,48 and Sz\,91 for which we consider respectively VISIR and Subaru observations, all have scattered light observations with~VLT/SPHERE.
We report the outer radius of the scattered light cavity and millimeter peak in~Table~\ref{tab:comparison}. We re-scale the published values using the latest distances from Gaia DR2 \citep{gaia_DR2_2018}. In this analysis, we only consider the position of the main cavity, regardless of the presence of an inner disk within the first few au.  For 5 of the 22 disks considered in this analysis, the scattered light cavity may be located inside the coronagraphic mask radius, giving the upper limits in Table~3. For the small fraction of transition disks that possesses multiple rings in scattered light, we report the position of the scattered light ring that is the closest to the millimeter peak emission (for the following objects: HD\,169142, V4046\,Sgr, J1852, HD\,97048, Lk\,Ca\,15, Sz\,91 and HD\,34282).
For J1608, which has a large inclination, we chose to estimate the ALMA and SPHERE peaks on our model, after computing it face-on. 

From synthetic observations, calculated after hydrodynamical and dust evolution simulations and considering massive planets on a circular orbit, \citet{Ovelar_2013} found that as the mass of a planet in a disk increases, the position of the millimeter ring moves further away from the planet's orbit while the outer radius of the scattered light cavity does not. They derived an analytic formula relating the planet mass with the ratio between the position of the so-called 'scattered light wall' to that of the ALMA peak. The scattered light wall is defined as the radial location where the scattered light signal is half of the difference between the flux measured at the peak of the wall and the minimum flux in the gap.  As the position of the wall is usually not explicitly published in the literature, we use both the inner radius of the disk beyond the cavity, as seen in scattered light, and the position of the peak in polarized intensity (respectively referred to as R$_{\rm{in,PI}}$, and R$_{\rm{peak, PI}}$, in Table\,\ref{tab:comparison}). Considering the position of the cavity instead of that of the wall tends to over-estimate the planet mass, and inversely under-estimate it when the position of the peak is used.  We also note that the models were specifically calculated for R band scattered light observations and Band\,7~(850\,$\mu$m), and only for a planet in a circular orbit. However, little difference is expected for such small variations in wavelength, as can be seen by comparing the theoretical profiles of Band\,R and Band\,H in Figure 3 of \citet{Ovelar_2013}.

We show in Fig.~\ref{fig:comparison_sizes} the radius of the scattered light cavity (R$_{\rm{cav,PI}}$),  defined as the mean between R$_{\rm{in,PI}}$ and~R$_{\rm{peak,PI}}$, as a function of the radius of the millimeter ring (R$_{\rm{peak,mm}}$).  We observe that for each system the scattered light cavity radius is smaller than the millimeter radius (green line), with about one third of the disks having a ratio smaller than~0.5 (see Table~\ref{tab:comparison}). 
The models of \citet{Ovelar_2013} would imply companion masses above 13\,M$_\mathrm{J}$ for ratios lower than 0.48 when the planet is located at 20\,au,~0.53 when it is at 40\,au, and 0.56 at~60\,au. 
As can be seen on Fig.~\ref{fig:comparison_sizes}, fifteen disks in our sample~(namely HD\,100546, HD\,169142, V4046\,Sgr, HD\,100453, T\,Cha, J1852, HD\,97048, Lk\,Ca\,15, Oph\,IRS\,48, PDS\,70, J1608, RXJ1604.3-2130A, Sz\,91, HD\,34282 and HD\,142527) are above the red shaded area. This indicates ratios larger than the ones given above, placing the possible companions in the planetary mass regime. PDS\,70 is the only system where a few Jupiter mass planet was imaged in the main cavity~\citep{Keppler_2018, Muller_2018}, while HD\,142527 has a stellar companion. For the other disks considered (namely UX\,Tau\,A, HD\,143006, DoAr44, CQ\,Tau, HD\,135344B, RY\,Lup and MWC758), the ratio~(or its upper limit) would lead to objects in the stellar or brown dwarf regime. These disks appear in the red shaded area in~Fig.~\ref{fig:comparison_sizes}.

Several direct imaging surveys have been carried out searching for companions. \citet{Kraus_2008, Kraus_2011} performed a high-resolution imaging studies of Taurus-Auriga and Upper Sco star-forming regions to identify companions down to 8 to~12\,M$_\mathrm{J}$. More recently, Subaru high-contrast observations of 68 young stellar objects were performed~\citep[SEEDS survey;][]{SEEDS_survey_2017}, reaching typical limits of~10\,M$_\mathrm{J}$ at~0.5\arcsec~($\sim$ 70\,au at 140\,pc) and 6\,M$_\mathrm{J}$ at 1\arcsec. The SEEDS survey covered 12 disks of our sample\footnote{Namely UX\,Tau\,A, HD\,143006, DoAr44, CQ Tau, J1852, HD\,135344B, Lk\,Ca\,15, Oph IRS 48, MWC 758, RXJ1604.3-2130A, Sz 91, HD\,34282}, without a planet detection. For HD\,169142, T\,Cha, HD\,135344B, HD\,97048 and RY\,Lup, detection limits were also presented in individual studies, that reached sensitivity of $\sim$\,10\,M$_\mathrm{J}$ at~0.25\arcsec \citep{Ligi_2018, Pohl_2017, Maire_2017, Ginski_2016, Langlois_2018}.  We also note that claims of candidate companions were made in the disks of HD\,100546~\citep{Quanz_2015, Currie_2014}, HD\,169142 \citep{Biller_2014, Reggiani_2014}, Lk\,Ca\,15 \citep{Kraus_2012, Sallum_2015} and MWC758 \citep{Reggiani_2018}, without having been firmly confirmed until now. 

An alternative scenario to explain the small ratio between the radius of the scattered light cavity and the millimeter peak could be that the cavities are caused by several lower mass planets, which would allow small grains to fill the cavity, while large grains are retained in the outer disk~\citep{Rosotti_2016, Dipierro_Laibe_2017}. In this case, no gap would be detected in scattered light, while a clear ring would appear in the millimeter images. 

\section{Conclusion}
\label{sec:conclusion}
In this paper, we present new polarized scattered light observations of two transition disks, namely 2MASS~J16083070-3828268~(J1608) and RXJ1852.3-3700~(J1852).
The image of J1608 reveals a highly inclined disk~($\sim$74$^\circ$) with a large cavity of about 50\,au in scattered light. We also detect a faint line, in the southwest, that we interpret as tracing the rear-facing side of the disk. The second disk of our study, J1852, shows scattered light (referred to as an inner ring) just beyond the coronagraph radius, a gap between 22 and~42\,au and an outer ring up to 65\,au. A cavity inward of the first ring, as inferred from the SED, is located behind the coronagraph. 

We modeled both scattered light and millimeter  images~(that trace small and large dust grains, respectively), together with the SED, using radiative transfer. Our modeling of the highly inclined disk J1608 indicates that small and large grains have a different spatial distribution. Radially, small grains are more extended inward and outward than the large grains, by respectively 30\,au and 60\,au. Vertically, at a radius of 100\,au, we constrain the large grains to be located within a height of 5\,au, while the small grains extend vertically up to 12\,au. We follow a similar procedure for~J1852 and propose a model with a spatial segregation between grain sizes. However, the disk is not inclined enough to allow us to strongly constrain the relative vertical extents of various grain sizes, and so our modeling of the images and SED for this object remains degenerate.

The radial and vertical segregation in particle sizes observed in J1608 is likely a consequence of both vertical settling and dust radial drift that occur during the evolution of the disks.  Vertical settling in low turbulence disks and/or following non-ideal MHD effects such as ambipolar diffusion can explain the relatively small scale height inferred for the large grain population.  The difference in the outer extents~(as measured in scattered light and millimeter emission) could result from radial drift, optical depth  and illumination effects, while the difference in the inner radius of the outer disk, might be related to the presence of planet(s). 

We compile a sample of 22 transition disks imaged with both ALMA and SPHERE, and find that scattered light is detected inside the millimeter cavity in all of the disks. We use the observed spatial difference in mm and far-IR distributions to identify a segregation in particle sizes,   and infer the proposed companion mass responsible for the cavity using the prescription of \citet{Ovelar_2013}. We show that in 15 objects, including the two disks modeled in this study, the cavities could be explained by the presence of a giant planet. The seven other disks of the sample show large ratios between the position of the scattered light and the millimeter cavity, suggestive of a companion above the planetary mass regime, or alternatively, of a multiple planetary system. 

As of today, apart from PDS\,70 \citep{Keppler_2018}, direct imaging surveys with results available in the literature, did not provide the detection of other such objects within a transition disk. New deeper observations with direct imaging instruments, or search for non-Keplerian motions in the gas kinematics with ALMA \citep{perezseba2015,perezseba2018, Teague_2018, pinte_2018} might lead to further detections. 

\begin{acknowledgements}
MV, MB, FM, GvdP, CP acknowledge funding from ANR of France under contract number ANR-16-CE31-0013 (Planet Forming Disks). M.F. received funding from the European Research Council (ERC) under the European Union's Horizon 2020 research and innovation programme (grant agreement n$^\circ$ 757957). J.\,O. acknowledges financial support from the ICM (Iniciativa Cient\'ifica Milenio) via the N\'ucleo Milenio de Formaci\'on Planetaria grant, from the Universidad de Valpara\'iso, and from Fondecyt (grant 1180395). L.P. acknowledges support from CONICYT project Basal AFB-170002 and from FONDECYT Iniciaci\'on project \#11181068. A.Z. acknowledges support from the CONICYT + PAI/ Convocatoria nacional subvenci\'on a la instalaci\'on en la academia, convocatoria 2017 + Folio PAI77170087.  This paper makes use of the following ALMA data: ADS/JAO.ALMA\#2015.1.00222.S and ADS/JAO.ALMA\#2015.1.01083.S. ALMA is a partnership of ESO (representing its member states), NSF (USA) and NINS (Japan), together with NRC (Canada), MOST and ASIAA (Taiwan), and KASI (Republic of Korea), in cooperation with the Republic of Chile. The Joint ALMA Observatory is operated by ESO, AUI/NRAO and NAOJ. This publication makes use of VOSA, developed under the Spanish Virtual Observatory project supported from the Spanish MINECO through grant AyA2017-84089.
SPHERE is an instrument designed and built by a consortium consisting of IPAG (Grenoble, France), MPIA (Heidelberg, Germany), LAM (Marseille, France), LESIA (Paris, France), Laboratoire Lagrange (Nice, France), INAF–Osservatorio di Padova (Italy), Observatoire de Genève (Switzerland), ETH Zurich (Switzerland), NOVA (Netherlands), ONERA (France) and ASTRON (Netherlands) in collaboration with
ESO. SPHERE was funded by ESO, with additional contributions from CNRS
(France), MPIA (Germany), INAF (Italy), FINES (Switzerland) and NOVA
(Netherlands).  SPHERE also received funding from the European
Commission Sixth and Seventh Framework Programmes as part of the
Optical Infrared Coordination Network for Astronomy (OPTICON) under
grant number RII3-Ct-2004-001566 for FP6 (2004–2008), grant number
226604 for FP7 (2009–2012) and grant number 312430 for FP7 (2013–2016). We also acknowledge financial support from the Programme National de
Planétologie (PNP) and the Programme National de Physique Stellaire
(PNPS) of CNRS-INSU in France. This work has also been supported by a grant from
the French Labex OSUG@2020 (Investissements d’avenir – ANR10 LABX56).
The project is supported by CNRS, by the Agence Nationale de la Recherche (ANR-14-CE33-0018). It has also been carried out within the frame of the National Centre for Competence in Research PlanetS supported by the Swiss National Science Foundation (SNSF). MRM, HMS, and SD are pleased to acknowledge this financial support of the SNSF. Finally, this work has made use of the the SPHERE
Data Centre, jointly operated by OSUG/IPAG (Grenoble), PYTHEAS/LAM/CESAM (Marseille), OCA/Lagrange (Nice) and Observtoire de
Paris/LESIA (Paris).
\end{acknowledgements}

\bibliographystyle{aa}
\bibliography{biblio}
\begin{appendix}
\section{Additional maps}
\label{sec:Qphi_Uphi}

\begin{figure*}
    \centering
    \includegraphics[width = 18cm]{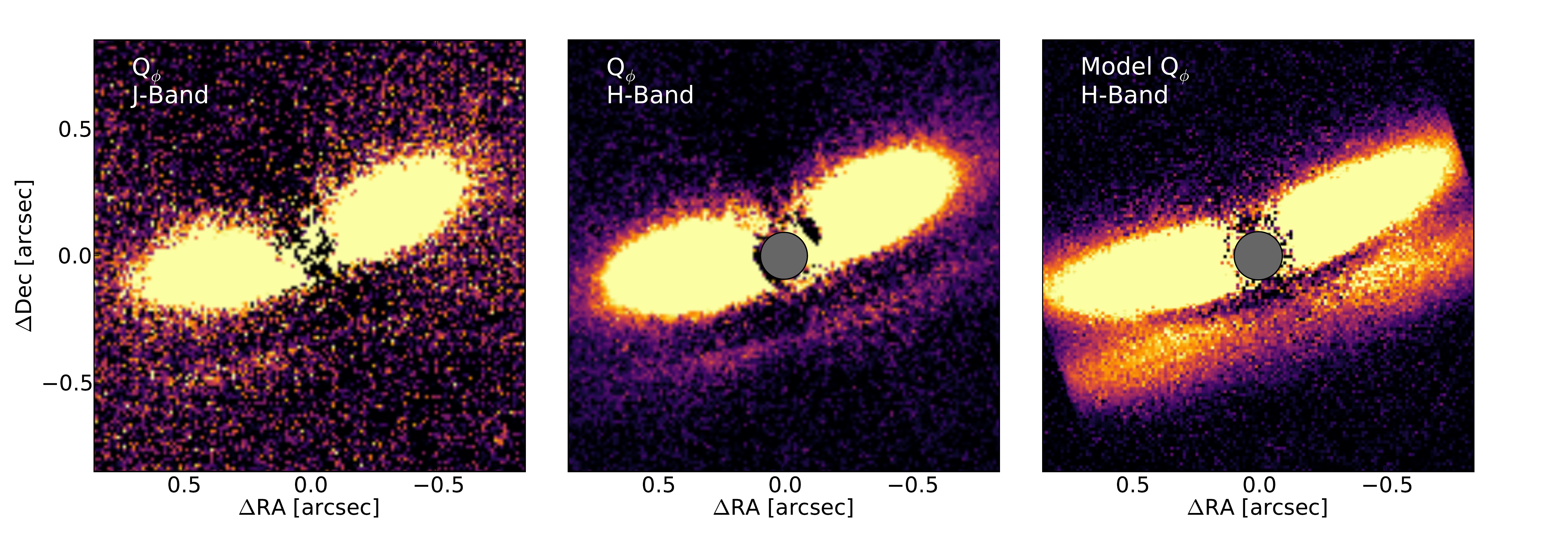}
    \caption{The Q$_\phi$ images of J1608 in J (non coronagraphic) and H-Band (coronagraphic) with a dynamical range from 0 to 5\% of the maximum of each image are showed in the left and middle panels, respectively. The bottom line of the disk is seen more clearly in the data than in Fig.~\ref{fig:data_J1608}, and appears to be too bright in the model (right panel).}
    \label{fig:J1608_faintline}
\end{figure*}

\paragraph{Saturated J1608 Q$_\phi$ image.} To visualize better the faint line south of J1608, we show the Q$_\phi$ image of the H and J-band observations in Fig~\ref{fig:J1608_faintline}, while saturating the image for brightness larger than 5\% of the maximum intensity of the map. The right panel also shows our Q$_\phi$ model image with the same dynamical range.

\begin{figure*}
    \centering
    \includegraphics[width = 18cm]{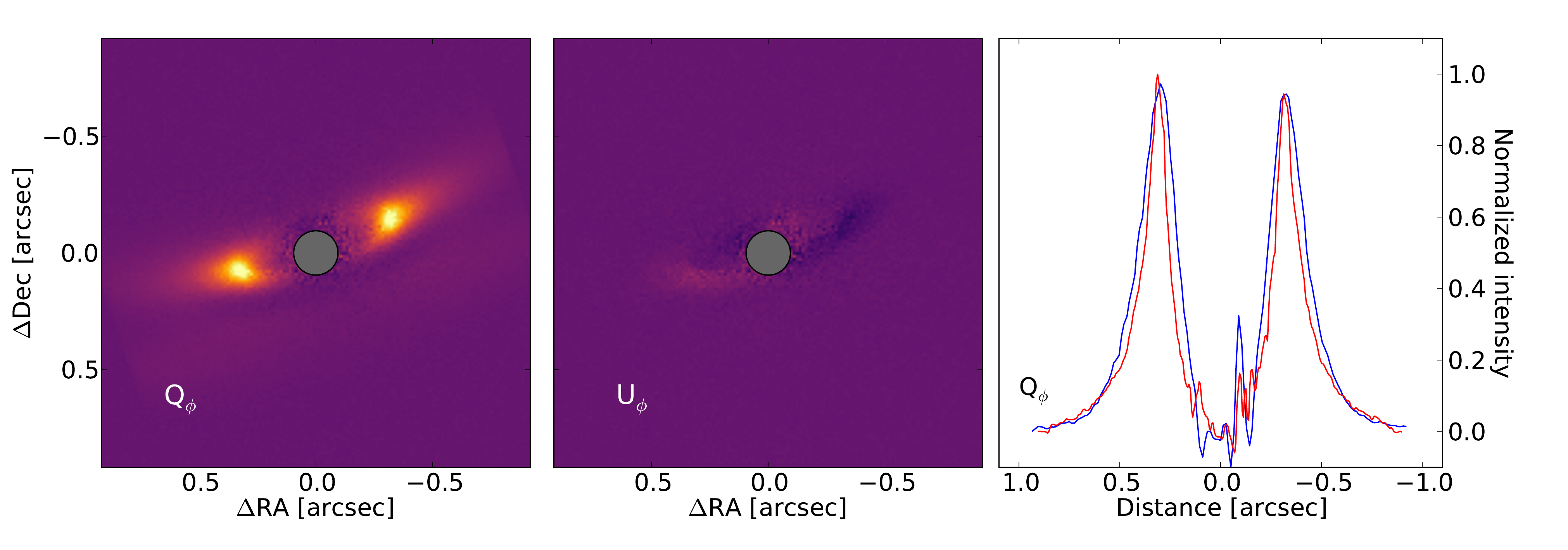}
    \caption{Left and middle: Q$_\phi$ and U$_\phi$ images of our model of J1608, respectively. The colorscale used in the same as in Fig.~\ref{fig:model_J1608}. The right panel shows radial cuts along the major axis, compared to the data.}
    \label{fig:J1608_Qphi_Uphi}
\end{figure*}

\begin{figure*}
    \centering
    \includegraphics[width = 18cm]{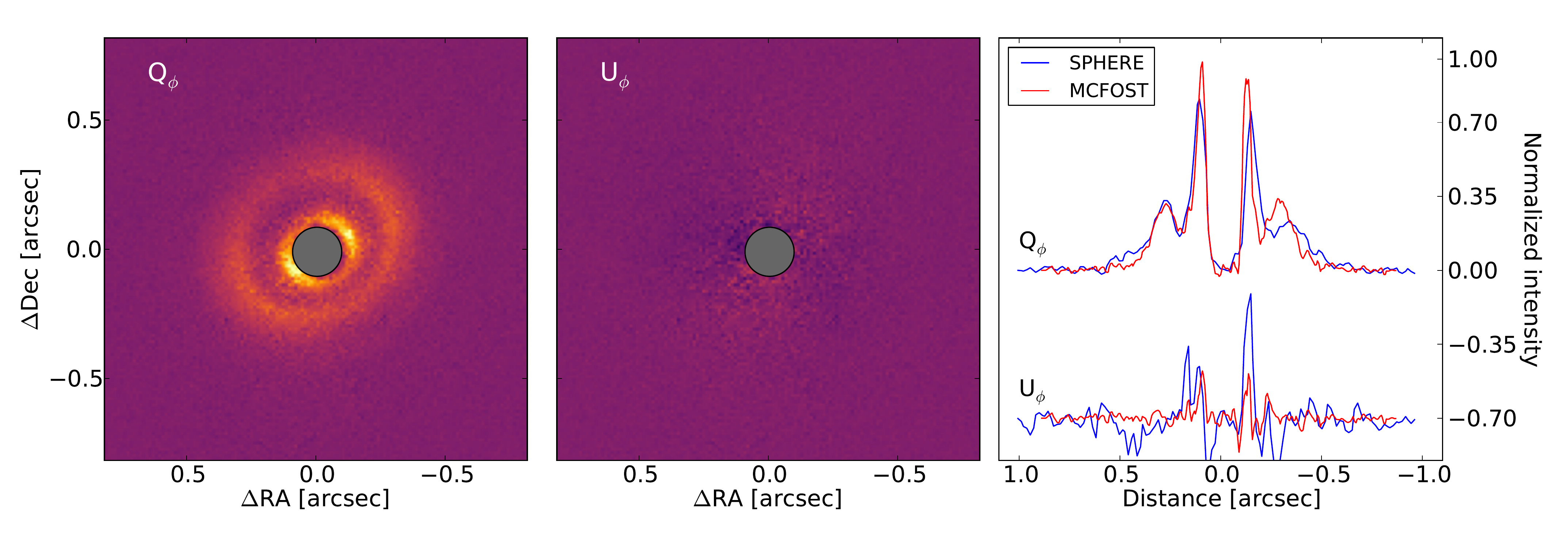}
    \caption{Left and middle: Q$_\phi$ and U$_\phi$ images of our model of J1852, respectively. The colorscale used in the same as in Fig.~\ref{fig:model_J1852}. The right panel shows radial cuts along the major axis, compared to the data.}
    \label{fig:J1852_Qphi_Uphi}
\end{figure*}

\paragraph{Model Q$_\phi$ and U$_\phi$ images.} As we show and model the polarized intensity  image in the main text, we present in Fig.~\ref{fig:J1608_Qphi_Uphi} and Fig.~\ref{fig:J1852_Qphi_Uphi}, the individual Q$_\phi$ and U$_\phi$ model predictions for J1608 and J1852, respectively. The Q$_\phi$ images are very similar to our model polarized intensity images that reproduce well the data. In addition, when compared to Fig.~\ref{fig:data_J1608} and Fig.~\ref{fig:data_J1852}, we see that the regions with positive and negative intensity in the U$_\phi$ images are also a good match to the data.

\begin{figure*}
     \centering
     \includegraphics[width = 18cm]{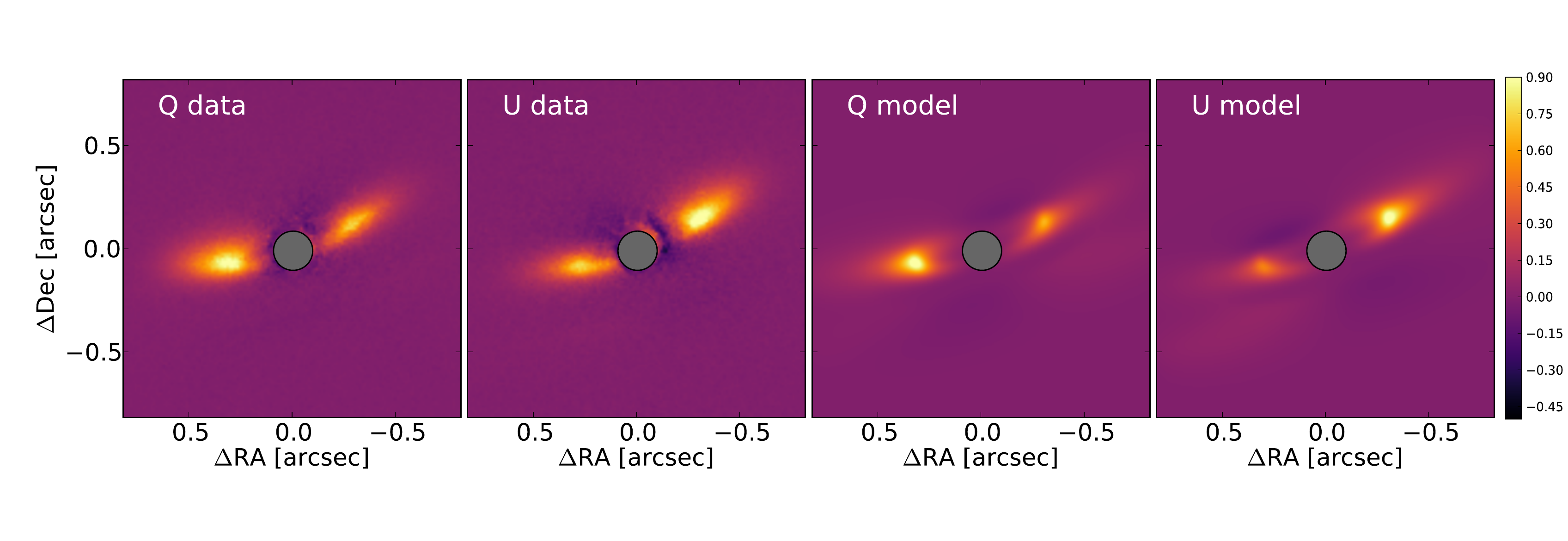}
     \caption{J1608 Q and U normalized maps of the data and model without noise.}
     \label{fig:J1608_Q_U}
 \end{figure*}

\begin{figure*}
     \centering
     \includegraphics[width = 18cm]{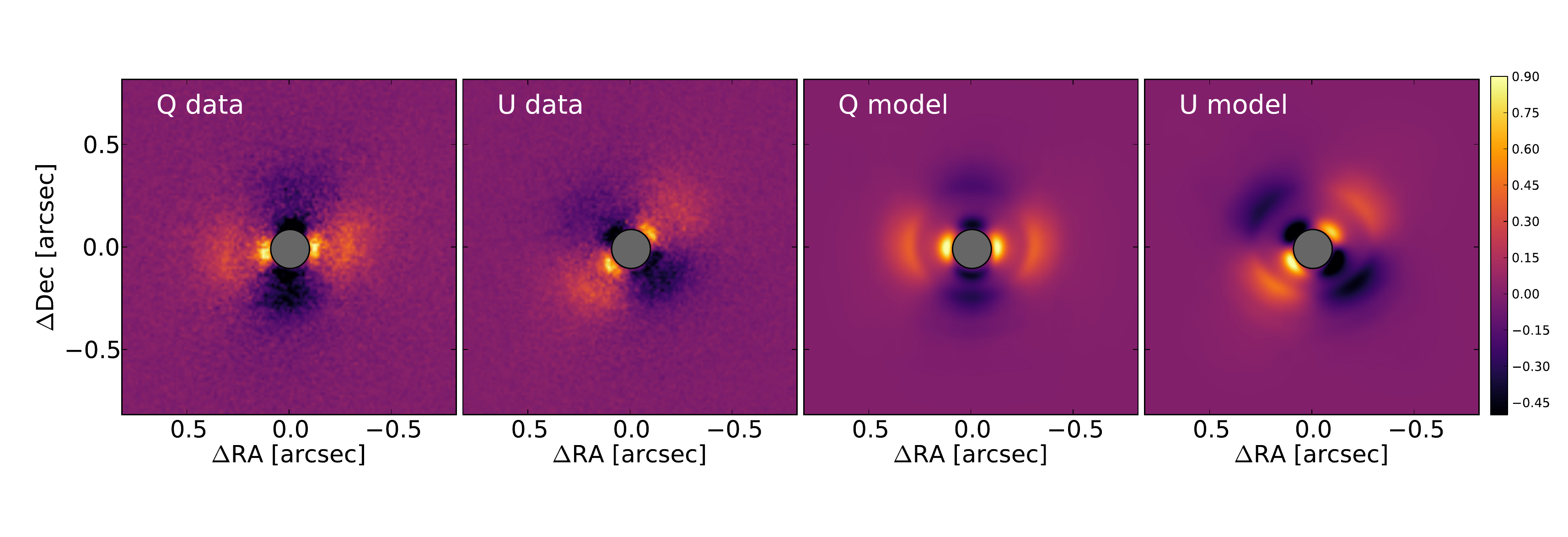}
     \caption{J1852 Q and U normalized maps of the data and model without noise.}
     \label{fig:J1852_Q_U}
 \end{figure*}

 \paragraph{Model Q and U images.} 
 We show the Q and U maps of both data and models~(without any noise) in Fig.~\ref{fig:J1608_Q_U} and~Fig.~\ref{fig:J1852_Q_U}. In J1608, the east/west asymmetry in the lobe is reproduced, and in J1852 we clearly see the two rings in the model images.

\begin{figure*}
    \centering
    \includegraphics[width = 14cm]{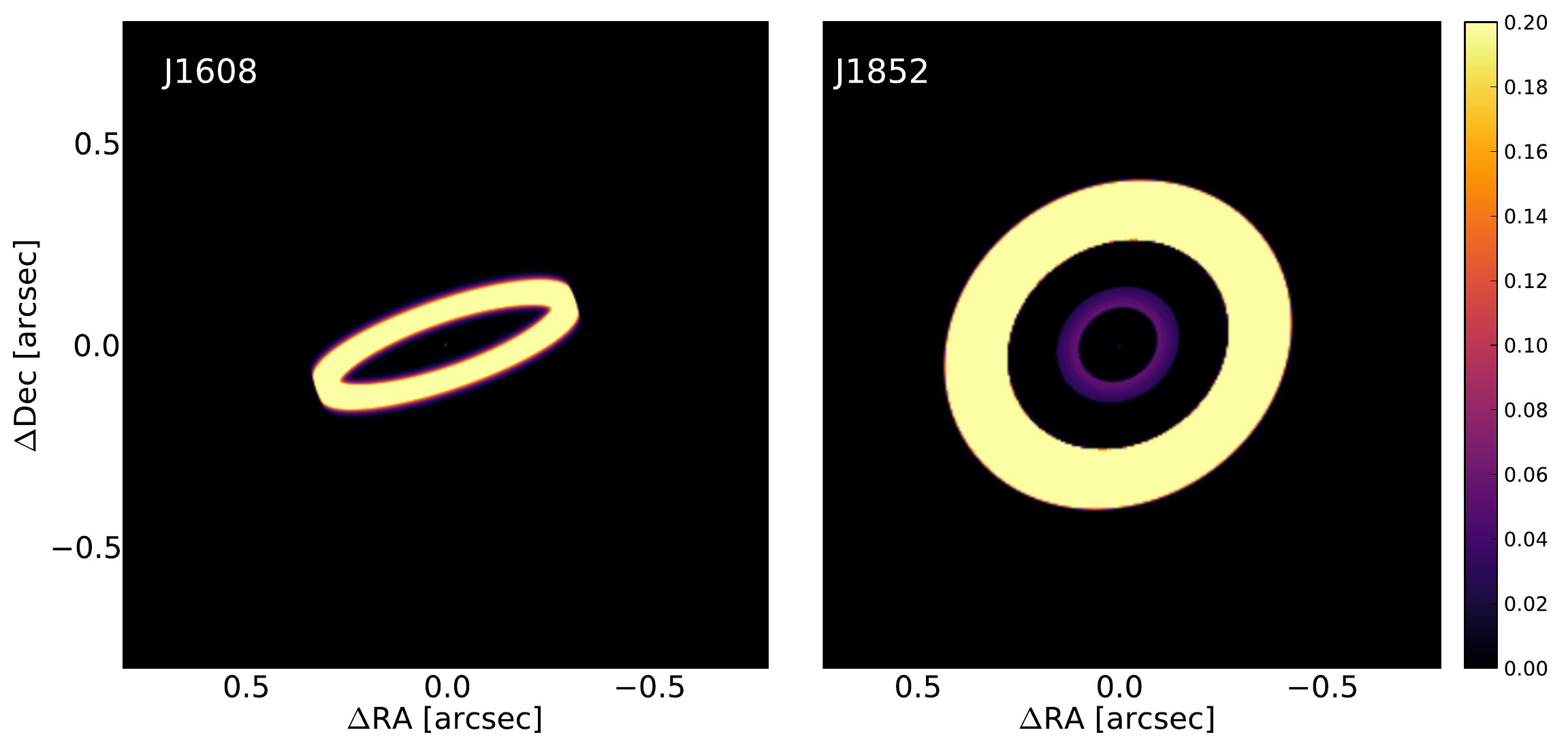}
    \caption{Synthetic millimeter predictions of our models for J1608 in Band 6 (left panel) and J1852 in Band 3 (right panel), before convolution by the ALMA beam.}
    \label{fig:Before_convolution}
\end{figure*}

\paragraph{Model millimeter images before convolution by the beam.} 
\label{sec:no_convolution}
In~Fig.~\ref{fig:Before_convolution}, we show our synthetic millimeter predictions for J1608 and J1852 before convolution by the corresponding ALMA beam.

\section{Model schematic representation}
\label{sec:modelsketch}
\begin{figure}[h]
    \centering
     \resizebox{\hsize}{!}{\includegraphics{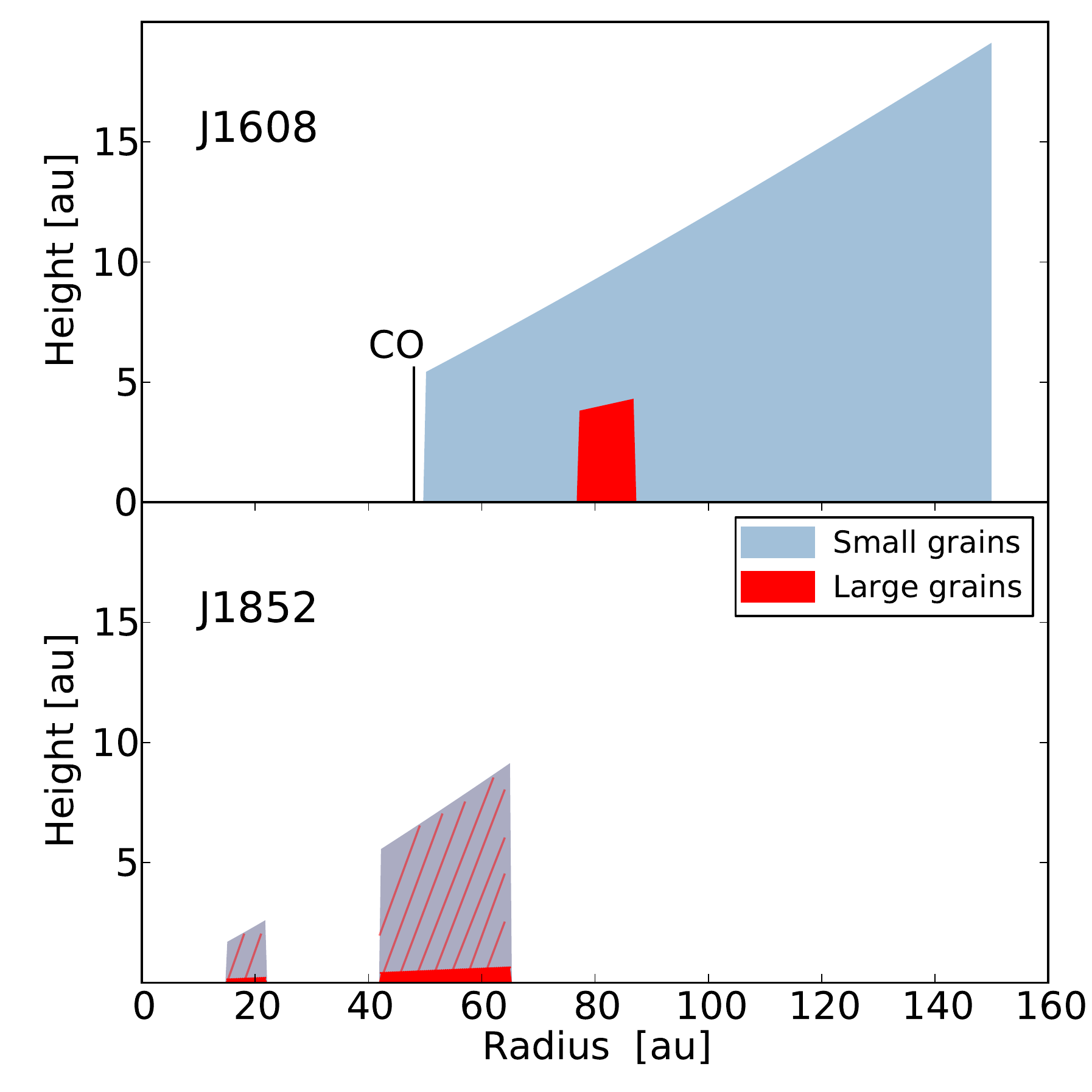}}
    \caption{Modeled radial and vertical structure for small and large grains in J1608 and J1852, in linear scale. The vertical black line indicates the inner radius of the disk in CO, as measured on the PV-diagram (Fig.~\ref{fig:diagramme_PV}). The red hatches in J1852 represent the uncertaincy on the scale height of the large grain population that is not well constrained by our model. The innermost disk of J1852, located between 0.2 and 2\,au, is too small to be visible in this representation.} 
    \label{fig:scale_heights}
\end{figure}

We show a schematic representation of our models~in~Fig.~\ref{fig:scale_heights}, that shows the clear spatial segregation in particle sizes for J1608. As explained in the main text, our model of J1852 remains degenerate. With a grain size distribution as given in Table~\ref{tab:parameters}, that uses a minimum grain size of 10$\mu$m for the large grain population, we obtain a good model represented in the bottom panel of Fig.~\ref{fig:scale_heights}. However, with a minimum grain size to 300\,$\mu$m, both small and large grain population could be mixed up to the same height. We therefore represent the height of the small grain population with red hatches to show this uncertainty.  

\end{appendix}
\end{document}